%% file: main_v12.tex
\documentclass[aps,10pt,onecolumn,superscriptaddress,nofootinbib,longbibliography]{revtex4-2}
\usepackage{amsmath,amssymb,amsthm}
\usepackage{xcolor,enumitem,graphicx}
\newcommand{\ket}[1]{\lvert#1\rangle}
\newcommand{\bra}[1]{\langle#1\rvert}
\usepackage[hidelinks]{hyperref}
\usepackage[nameinlink,noabbrev]{cleveref}
\crefname{equation}{Eq.}{Eqs.}\Crefname{equation}{Eq.}{Eqs.}
\crefname{section}{Sec.}{Secs.}\Crefname{section}{Sec.}{Secs.}
\crefname{subsection}{Sec.}{Secs.}\Crefname{subsection}{Sec.}{Secs.}
\crefname{subsubsection}{Sec.}{Secs.}\Crefname{subsubsection}{Sec.}{Secs.}
\crefname{appendix}{Appendix}{Appendices}\Crefname{appendix}{Appendix}{Appendices}
\crefname{figure}{Fig.}{Figs.}\Crefname{figure}{Fig.}{Figs.}
\crefname{Thm}{Theorem}{Theorems}\Crefname{Thm}{Theorem}{Theorems}
\crefname{assumption}{Assumption}{Assumptions}
\Crefname{assumption}{Assumption}{Assumptions}
\crefname{enumi}{Step}{Steps}\Crefname{enumi}{Step}{Steps}
\newtheorem{Thm}{Theorem}
\newtheorem{Lem}{Lemma}
\crefname{Lem}{Lemma}{Lemmas}\Crefname{Lem}{Lemma}{Lemmas}
\usepackage{tikz,pgfplots}
\usepackage{physics}
\pgfplotsset{compat=1.18}

\input{notation_final}
\renewcommand{\conditionalMain}{p_{\rm main}}
\renewcommand{\conditionalJoint}{\widetilde p_{\rm main}}
\renewcommand{\conditionalTarget}{p_{1_m}}
\newcommand{\conditionalMainMu}{p_{{\rm main}|\meanPhoton}}
\newcommand{\conditionalJointMu}{\widetilde p_{{\rm main}|\meanPhoton}}
\newcommand{\conditionalTargetExact}{p_{1_m}^{\rm exact}}
\renewcommand{\virtualBranch}{C} 
\renewcommand{\darkFreeWeight}{u} 
\renewcommand{\inputJointEffect}{\hat J_{el}^{0}} 

\begin{document}

\title{Decoy-state quantum key distribution with a leaky Mach--Zehnder receiver}

\author{Ryohei Mizusawa}

\affiliation{Faculty of Engineering, University of Toyama, Gofuku 3190, Toyama 930-8555, Japan}

\author{Álvaro Navarrete}

\affiliation{Faculty of Engineering, University of Toyama, Gofuku 3190, Toyama 930-8555, Japan}

\affiliation{Vigo Quantum Communication Center, University of Vigo, Vigo E-36310, Spain}

\affiliation{School of Telecommunication Engineering, Department of Signal Theory and Communications, University of Vigo, Vigo E-36310, Spain}

\affiliation{atlanTTic Research Center, University of Vigo, Vigo E-36310, Spain}

\author{Akihiro Mizutani}

\affiliation{Faculty of Engineering, University of Toyama, Gofuku 3190, Toyama 930-8555, Japan}

\author{Kiyoshi Tamaki}

\affiliation{Faculty of Engineering, University of Toyama, Gofuku 3190, Toyama 930-8555, Japan}

\begin{abstract}
We prove the security of phase-encoded decoy-state BB84 quantum key distribution for a Mach--Zehnder receiver with imbalanced beam splitters and two threshold detectors that leak fine-grained information about the detection patterns. For this, we derive a lower bound on Bob's single-photon contributions and combine it with the decoy-state method and a recent phase-error framework for leaky receivers. The leakage-induced correction to the phase-error rate is independent of channel loss, providing a loss-tolerant treatment of receiver leakage even in the presence of beam-splitter imbalance. Our results provide a practical route toward loss-tolerant security proofs for imperfect QKD receivers in the presence of receiver leakage.
\end{abstract}

\maketitle

\section{Introduction}
\label{sec:intro}

Quantum key distribution (QKD) allows two users, Alice and Bob, to establish a secret key over an untrusted quantum channel, provided that their devices satisfy the assumptions of the security proof~\cite{bb84,lo2014secure,pirandola2020advances,xu2020secure}. In practice, however, deviations from the assumed source and receiver models may give an eavesdropper, Eve, advantages not accounted for in the security proof and thereby undermine its guarantees~\cite{marquardt2024implementation,zapatero2025}. Incorporating these deviations into the security analysis is therefore crucial to establish the security of practical implementations.

On the source side, device imperfections such as state-preparation flaws, side-channel leakage, and correlated emissions have been extensively addressed, either separately or in combination~\cite{Gottesman2004ImperfectDevices,Tamaki2014LossTolerant,Pereira2020CorrelatedSources,CurrasLorenzo2025ImperfectSources,curras2026phase,curras2025numerical,Navarrete2026NumericalPracticalQKD,wiesemann2026security}.
On the receiver side, measurement-device-independent (MDI) QKD removes the need to trust the measurement apparatus altogether~\cite{Lo2012MDIQKD}. However, conventional prepare-and-measure schemes, which are commonly adopted in commercial systems, typically require an explicit model of Bob's receiver. To simplify this task, the so-called qubit squashing models~\cite{TsurumaruTamaki2008ThresholdDetectors,Beaudry2008Squashing,Gittsovich2014Squashing} reproduce the statistics of certain receivers with an effective qubit measurement. Flag-state squashing~\cite{Zhang2021DetectionEfficiencyMismatch}, in contrast, preserves the actual measurement up to a photon-number cutoff and replaces the higher-photon-number contributions with classical flags, which requires a lower bound on the weight of the preserved subspace. Such bounds have been obtained from cross-click statistics in phase-encoded BB84 with unbalanced signal amplitudes~\cite{Li2020FlagStateSquashing} and, analytically, from multiclick statistics in general passive linear-optical receivers~\cite{Kamin2024ImprovedFlagState}. Alternatively, detector-decoy methods estimate the photon-number statistics of the incoming signals by means of a variable attenuator~\cite{moroder2009detector}. 
Partially characterized detector efficiencies and dark counts have also been incorporated into security proofs~\cite{tupkary2025phase,CurrasLorenzo2026SourceDetector,Nahar2026ImperfectDetectors}.

Another important type of imperfection at the receiver is information leakage---e.g., through backflash radiation~\cite{Kurtsiefer2001Backflash,Meda2017Backflash,Pinheiro2018Backflash}. The framework of Ref.~\cite{leakyreceiver2026} accommodates such leakage in a loss-tolerant manner---i.e., the effect of the leakage is not amplified by channel loss, unlike in existing techniques~\cite{maroySecurityQuantum2010,arqandMutualInformation2024}---given that a bound on the fidelity between two leakage states is known. Applying it to a specific optical receiver, however, requires an effective qubit description of the measurement compatible with the recorded bit values and the leaked information. Standard qubit squashing models are not suited for this purpose. This is because they replace the optical measurement by an effective qubit measurement that reproduces only the statistics of the recorded outcomes, whereas the leakage may depend on the actual underlying physical detection pattern---e.g., on whether an event was a single or a double click, the latter of which may be assigned a uniformly random bit---which has no counterpart in the squashed picture. 

In this work, we address this type of imperfection for phase-encoded decoy-state BB84 with a receiver consisting of an asymmetric Mach--Zehnder interferometer (AMZI) and two threshold detectors, allowing for a partially characterized entrance beam-splitter transmittance, an uncharacterized exit transmittance, and partially characterized dark-count probabilities. In this receiver, each pulse pair has a main detection window, in which its two pulses interfere, flanked by two satellite windows shared with the neighboring pairs. To accommodate the leakage, we restrict the analysis to the single-photon component in the main detection modes, and derive an operator inequality that allows us to lower bound the probability that these modes contain exactly one photon in terms of the click statistics in the main and satellite windows. Combining this bound with the receiver-leakage analysis of Ref.~\cite{leakyreceiver2026} and the decoy-state method~\cite{Lo2005DecoyState,charles2013}, we obtain an analytical lower bound on the asymptotic secret-key rate that accounts explicitly for the leakage from the detection patterns. The impact of the leakage on the phase-error rate is loss-tolerant, i.e., independent of the channel loss, even for imbalanced beam splitters.

\section{Setup and assumptions}
\label{sec:main}

\begin{figure}[tb]
\centering
\includegraphics[width=0.96\textwidth]{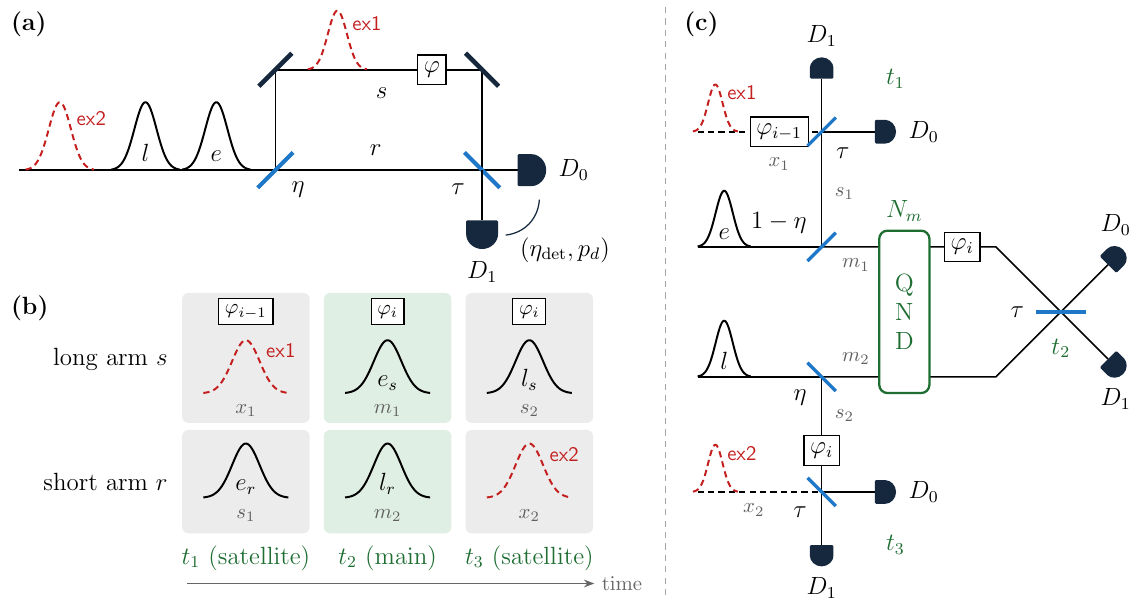}
\caption{(a) Bob's receiver setup. The long and short arms are labeled $s$ and $r$, respectively. In the notation used below, $\mainMode{1}=e_s$, $\mainMode{2}=l_r$, $\satMode{1}=e_r$, and $\satMode{2}=l_s$; the pulses labeled ex1 and ex2 supply $\extraMode{1}$ and $\extraMode{2}$. The window $t_2$ is the main window, while $t_1$ and $t_3$ are satellite windows shared with the preceding and following pairs. (b) Correspondence between the pulses in each arm and the detection windows. For pair $i$, the long-arm phase modulator applies $\bobPhase_i$ to both the early and late pulses. The long-arm phase settings in windows $t_1,t_2,t_3$ are therefore $\bobPhase_{i-1},\bobPhase_i,\bobPhase_i$, respectively. 
(c) Equivalent optical layout of the main and satellite modes, including a virtual quantum non-demolition (QND) measurement of the total photon number $N_m$ in the main modes $m_1$ and $m_2$.}
\label{fig:receiver}
\end{figure}

We consider phase-encoded decoy-state BB84. Alice sends early--late pulse pairs, and Bob measures them with the Mach--Zehnder interferometer in \cref{fig:receiver}. We first specify the source and the receiver model used in the security analysis.

\subsection{Source model}
\label{subsec:source}

Alice prepares pairs of coherent pulses with total mean photon number $\meanPhoton$ and relative phase $\signalPhase\in\{0,\pi/2,\pi,3\pi/2\}$. For a common phase $\globalPhase$, the two-pulse state is
\begin{equation}
  \ket{\sourceKetSymbol(\globalPhase,\signalPhase)}_{\signalSystem}
  :=
  \ket*{\sqrt{\meanPhoton/2}e^{i\globalPhase}}_{\earlySignalMode}
  \otimes
  \ket*{\sqrt{\meanPhoton/2}e^{i(\globalPhase+\signalPhase)}}_{\lateSignalMode}.
  \label{eq:pure_double_pulse}
\end{equation}
Here $\signalSystem:=\earlySignalMode\lateSignalMode$ denotes the two emitted temporal modes, and $\ket{\coherentAmplitude}$ denotes a coherent state with complex amplitude $\coherentAmplitude$. We define the normalized state with total photon number $n$ and relative phase $\signalPhase$ as
\begin{equation}
  \ket{n,\signalPhase}_{\signalSystem}
  :=
  \frac{1}{\sqrt{2^n}}
  \sum_{m=0}^{n}
  \sqrt{\binom{n}{m}}e^{im\signalPhase}
  \ket{n-m}_{\earlySignalMode}
  \otimes
  \ket{m}_{\lateSignalMode}.
  \label{eq:nphiS}
\end{equation}
The common phase $\globalPhase$ is uniformly randomized over $[0,2\pi)$ and is unknown to Eve. Averaging over this phase gives a mixture of states with different total photon numbers,
\begin{equation}
  \label{eq:appratus:state}
  \begin{aligned}
    \density_{\signalSystem}^{(\meanPhoton,\signalPhase)}
    &:=
    \frac{1}{2\pi}\int_0^{2\pi}
    \ket{\sourceKetSymbol(\globalPhase,\signalPhase)}_{\signalSystem}
    \bra{\sourceKetSymbol(\globalPhase,\signalPhase)}
    d\globalPhase \\
    &=
    \sum_{n=0}^{\infty}
    \photonProbability{n}
    \ket{n,\signalPhase}_{\signalSystem}
    \bra{n,\signalPhase},
  \end{aligned}
\end{equation}
where the photon-number probabilities follow the Poisson distribution
\begin{equation}
  \photonProbability{n}
  := e^{-\meanPhoton}\frac{\meanPhoton^n}{n!}.
  \label{eq:poisson}
\end{equation}
Note that the relative-phase coherence is retained within each fixed-$n$ component. Note also that, for fixed $n$ and $\signalPhase$, the state $\ket{n,\signalPhase}_{\signalSystem}$ is independent of the intensity label, which enters only through the Poisson weights $\photonProbability{n}$. This intensity independence is the source property used in the decoy-state analysis.

\subsection{Receiver model and scope}
\label{subsec:asm}

Bob's receiver is shown in \cref{fig:receiver}, and it consists of a Mach--Zehnder interferometer with entrance and exit beam-splitter transmittances $\entranceTrans$ and $\exitTrans$. Ideally, $\entranceTrans=\exitTrans=1/2$. However, for the security proof, we only need to assume that $\entranceTrans$ is known to lie within a certain calibration interval $[\entranceTrans_-,\entranceTrans_+]$, with $0<\entranceTrans_-\leq\entranceTrans_+<1$. As for $\exitTrans$, it can be completely uncharacterized. Moreover, a basis-independent loss in one arm of the interferometer---such as the insertion loss of the phase modulator~\cite{Li2020FlagStateSquashing}---is covered by this model. This is because an entrance beam splitter of transmittance $\entranceTrans$ followed by a transmittance $\kappa$ in the long arm is equivalent to a common loss of transmittance $\entranceTrans+(1-\entranceTrans)\kappa$ that can be attributed to the channel, followed by a lossless beam splitter of transmittance $\entranceTrans_{\rm eff}=\entranceTrans/[\entranceTrans+(1-\entranceTrans)\kappa]$. Calibration intervals for $\entranceTrans$ and $\kappa$ thus translate into an interval for $\entranceTrans_{\rm eff}$, which is all that the security analysis requires. 

We denote the incoming early and late modes of the current pulse pair by $\earlyInputMode$ and $\lateInputMode$, respectively, and we define the entrance isometry $\entranceIsometry$ as the map from their Fock space to that of the main modes $\mainMode{1}$ and $\mainMode{2}$ ($\mainSystem\equiv\mainMode{1}\mainMode{2}$) and the satellite modes $\satMode{1}$ and $\satMode{2}$ ($\satelliteSystem\equiv\satMode{1}\satMode{2}$) (see \cref{fig:receiver}c). A photon in mode $\earlyInputMode$ is directed to $\mainMode{1}$ with probability $1-\entranceTrans$ and to $\satMode{1}$ with probability $\entranceTrans$, while a photon in mode $\lateInputMode$ is directed to $\mainMode{2}$ with probability $\entranceTrans$ and to $\satMode{2}$ with probability $1-\entranceTrans$. The satellite modes interfere with certain extra modes, namely $\extraMode{1}$ and $\extraMode{2}$, whose state may be correlated with that in $\earlyInputMode$ and $\lateInputMode$, as well as with Eve's systems.

Let $\bobPhase_i$ denote the long-arm phase setting for pair $i$, measured relative to Bob's $Z$-basis setting. Bob sets $\bobPhase_i=0$ for the $Z$ basis and $\bobPhase_i=\pi/2$ for the $X$ basis. Here we assume that the long-arm phase modulator applies the same phase $\bobPhase_i$ to both pulses of pair $i$, namely to the main mode $\mainMode{1}=e_s$ and the satellite mode $\satMode{2}=l_s$. Consequently, the long-arm phase settings in windows $t_1,t_2,t_3$ of \cref{fig:receiver} are $\bobPhase_{i-1},\bobPhase_i,\bobPhase_i$, respectively\footnote{Indeed, we note that applying the same phase to both long-arm pulses is compatible with our security proof, but not required: the satellite windows enter the proof only through their no-click events, which are phase independent because both output ports are monitored, and through their leakage, which is given to Eve in full. With sufficiently fast modulation, Bob may instead phase-shift only $\mainMode{1}$, provided that \cref{asm:causal} still holds.}. In what follows, we label the satellite window shared by pairs $i$ and $i+1$---which combines mode $\satMode{2}$ of pair $i$ with mode $\satMode{1}$ of pair $i+1$, and thus uses the long-arm phase $\bobPhase_i$---by the half-integer index $i+1/2$. Its two-detector record is denoted by $\satRecord{i+1/2}\in\{00,10,01,11\}$, where $00$ denotes no clicks at either detector, $10$ ($01$) denotes a click only at $\detector{0}$ ($\detector{1}$), and $11$ denotes clicks at both detectors. The record of the main window is denoted by $\mainRecord\in\{00,10,01,11\}$.

We consider the following assumptions.

\begin{enumerate}[label=\textbf{A\arabic*:},ref=A\arabic*,leftmargin=*]

\item
\label[assumption]{asm:optical}
\textbf{Optical modes and detectors.}
We assume that only the optical modes specified above contribute to detection and that the unused entrance ports are in vacuum. Both detectors have the same efficiency---which is independent of the basis choice---in all modeled modes and windows. That is, each detector can be modeled as a linear loss followed by an ideal threshold detector.
Dark counts occur randomly and independently of each other, of the optical input, of the basis choices, and of Eve's pre-existing systems: in each physical window, each detector registers a dark count---i.e., a click that occurs irrespective of the optical input---with a probability $\darkProb$ common to both detectors. This probability is only known to lie in a calibration interval $[\darkProb^-,\darkProb^+]$, with $0\leq\darkProb^-\leq\darkProb^+<1$, and may vary between windows (see \cref{app:robust} for an extension to detectors with unequal, partially characterized dark-count probabilities). We assume no dead-time or afterpulsing effects.

\item
\label[assumption]{asm:causal}
\textbf{Protected target input and fresh basis choice.}

Bob's receiver is enclosed by a protected boundary, beyond which Eve has no access to its internal optical modes. For each pair, both target inputs $\earlyInputMode$ and $\lateInputMode$ cross this boundary before any leakage emitted in a detection window involving $\earlyInputMode$ or $\lateInputMode$ becomes available to Eve. Hence, Eve cannot use the leakage associated with the current pair to prepare or manipulate that pair, although she may use it to adapt the inputs of later pairs. Moreover, Bob chooses his basis for each pair freshly and independently, and no other side channel reveals it before the pair has crossed the boundary.

\item
\label[assumption]{asm:leak}
\textbf{Fresh pattern-conditioned leakage.}
At each main window the receiver produces the record $\mainRecord$ and freshly prepares a state $\leakState_{\mainLeakageSystem}^{(\mainRecord,w)}$, where $\mainLeakageSystem$ is the emitted leakage system. Similarly, at the satellite window $i+1/2$ it freshly prepares a state $\leakState_{\satLeakageSystem{i+1/2}}^{(\satRecord{i+1/2},\bobPhase_i)}$. These states have no additional dependence on the photon number, the input state, or past records. Conditioned on the current detection record and receiver setting, the newly prepared leakage state is in a tensor product with all pre-existing systems. A shared window has one record and one emission.

Moreover, we assume that, for the main single clicks in the key basis, there is a characterized constant $0\leq\leakOverlap\leq1$ satisfying
\begin{equation}\label{eq:Bobleaky}
 F\!\left(
 \leakState_{\mainLeakageSystem}^{(10,Z)},
 \leakState_{\mainLeakageSystem}^{(01,Z)}
 \right)\geq\leakOverlap^2,
 \qquad
 F(\density,\leakState):=
 \|\sqrt{\density}\sqrt{\leakState}\|_1^2,
\end{equation}
for any round, where $\|\cdot\|_1$ is the trace norm and $F$ is the fidelity. No overlap bound is required between the leakage states of double-click and single-click events, nor between the leakage states associated with the satellite windows.

\item
\label[assumption]{asm:source}
\textbf{Source and classical processing.}
Alice uses independent phase-randomized pairs as in \cref{subsec:source}, with independent private choices of bit, basis, and intensity. At fixed emitted photon number, the state is independent of the intensity label. As shown in \cref{subsec:virtual}, her bit-averaged single-photon state is then basis independent, and her complementary virtual measurement reproduces the test preparations. The users have trusted private randomness and an authenticated public channel, and do not announce current source labels during transmission. We consider the asymptotic limit of $\totalPairs$ transmitted pairs.

\end{enumerate}

Because the linear loss is the same for all detection modes under \cref{asm:optical}, it commutes with the beam splitters and phase shifts. We therefore move this common loss before the receiver---i.e., we give it to Eve---and subsequently use ideal threshold detectors with the dark counts specified in \cref{asm:optical}. Accordingly, all receiver-side photon numbers below are defined after this common loss. The case of detectors with unequal dark-count probabilities is addressed in \cref{app:robust}.

\Cref{asm:causal} allows us to describe the target pulse pair by an input state $\density_{\earlyInputMode\lateInputMode}$, fixed before Eve receives any leakage associated with that pair, followed by the trusted entrance isometry $\entranceIsometry$. 
This description is used in \cref{app:proof_lmm_qn} to derive the main-mode single-photon probability bound in Theorem~\ref{thm:single_bound}. The previous round's late satellite window $t_3$ coincides with the current round's early satellite window $t_1$, whose outcome also depends on the current early input $e$ through mode $s_1$. Leakage from this shared window must therefore remain unavailable to Eve until the current late input $\lateInputMode$ has crossed Bob's protected boundary. 
A sufficient condition to meet \Cref{asm:causal} is that the propagation time from the protected boundary to the earliest relevant detection via the short arm exceeds the early--late pulse separation: both inputs are then inside the boundary before detection occurs. Receivers satisfying this condition require no additional delay. Otherwise, an optical delay can be inserted inside the protected boundary, before the entrance beam splitter.

\section{Actual protocol}
\label{subsec:actual}

\begin{enumerate}[label=\textbf{\arabic*:},ref=\arabic*,leftmargin=*]

\item
\label{step:preparation}
\textbf{Parameters and state preparation.}
Alice and Bob fix the number $\totalPairs$ of transmitted pairs, the basis probabilities $\basisProbability{Z},\basisProbability{X}>0$, with $\basisProbability{Z}+\basisProbability{X}=1$, and a set of three intensities $\intensitySet:=\{\meanPhoton_1,\meanPhoton_2,\meanPhoton_3\}$ satisfying $\meanPhoton_1>\meanPhoton_2+\meanPhoton_3$ and $\meanPhoton_2>\meanPhoton_3\ge0$, together with their selection probabilities $\intensityProbability>0$. Then, for each pair, Alice independently selects a uniformly random bit $b$, a basis $w_{\aliceSystem}$ with probability $\basisProbability{w_{\aliceSystem}}$, and an intensity $\meanPhoton\in\intensitySet$ with probability $\intensityProbability$, and sends the state $\density_{\signalSystem}^{(\meanPhoton,\signalPhase)}$ in \cref{eq:appratus:state} with relative phase $\signalPhase=0$ ($\pi$) for $b=0$ ($1$) in the $Z$ basis and $\signalPhase=\pi/2$ ($3\pi/2$) for $b=0$ ($1$) in the $X$ basis. Bob independently chooses his basis $w_{\bobSystem}$ with the same probabilities and sets his long-arm phase accordingly, i.e., $\bobPhase_i=0$ for $Z$ and $\bobPhase_i=\pi/2$ for $X$ (see \cref{subsec:asm}).

\item
\label{step:detection}
\textbf{Detection and recording.}
Bob records the main pattern $\mainRecord$ and the two satellite records $\satRecord{i\pm1/2}$ of each pair. The main patterns $10$ and $01$ are assigned the bit values $0$ and $1$, respectively, while a double click is assigned a uniformly random bit drawn from Bob's trusted private randomness. After the last transmission he allows the last relevant detection windows to finish.

\item
\label{step:sifting}
\textbf{Sifting.}
Over an authenticated public channel, the users announce the bases, intensity labels, and main detection positions. The main clicks for which both users selected the $Z$ ($X$) basis define the key (test) rounds. In addition, for each basis $w$ and intensity $\meanPhoton$, Bob announces the number of main clicks with matching basis $w$ and intensity $\meanPhoton$ for which both neighboring satellite records are $00$; the satellite records are used for parameter estimation only.

\item
\label{step:estimation}
\textbf{Parameter estimation.}
The users reveal all test-round bits and count the bit errors for each intensity. In the asymptotic limit its size diverges but its fraction vanishes, so its removal does not change the limiting normalized gains. Then, they compute the target lower bounds $\targetLower^{Z},\targetLower^{X}$ in \cref{eq:Q1_lower_bound}, the test-error upper bound $\testErrorUpper$ in \cref{eq:E11_upper_bound}, and the phase-error rate bound $\entropyPhase$ in \cref{eq:eph}.

\item
\label{step:reconciliation}
\textbf{Reconciliation and privacy amplification.}
Alice's unrevealed bits from the remaining key rounds form the sifted key. The users perform error correction, verify agreement, and abort if verification fails. They account for the public reconciliation disclosure and apply universal$_2$ privacy amplification at the asymptotic rate of \cref{eq:key_rate_final}.

\end{enumerate}

\section{Security analysis}
\label{sec:app}

Here we bound the number and phase-error rate of the target events $(1_{\aliceSystem},1_m)$, in which Alice emits one photon and Bob's main modes contain one photon, with all receiver-side photon numbers defined after the common loss has been moved to the channel, as explained in \cref{subsec:asm}. All other events are treated as \emph{tagged}, i.e., we conservatively assume that Eve has full information about their bit values, so that they contribute to the sifted key but not to the secret key. The number and phase-error rate of the target events thus determine the privacy-amplification term in the key rate.

\subsection{Main photon number and preservation of leakage}
\label{subsec:modes}

For round $i$, let $\mainEvent^{(i)}$ denote the event ``at least one click in the main window'', and $\satVacEvent^{(i)}$ denote ``no click at either detector in either of its two neighboring satellite windows''. Using the satellite records $\satRecord{i\pm1/2}$ defined in \cref{subsec:asm}, we have
\begin{equation}\label{eq:shared-record}
 \satVacEvent^{(i)}:=\{\satRecord{i-1/2}=00,\ \satRecord{i+1/2}=00\}.
\end{equation}
Note that the record $\satRecord{i+1/2}$ enters both $\satVacEvent^{(i)}$ and $\satVacEvent^{(i+1)}$, since its window is shared by pairs $i$ and $i+1$. Each physical dark-count trial and leakage emission in a satellite window thus occurs only once, even though it affects the events of two neighboring pairs.

Now, define
\begin{equation}\label{eq:main-number}
 \mainNumberOp:=\numberOperator_{\mainMode{1}}+\numberOperator_{\mainMode{2}},\qquad
 \mainProjector{n}:=\indicator[\mainNumberOp=n],\qquad
 \numberChannel(\density):=\sum_{n\geq0}(\mainProjector{n}\otimes\II)\density(\mainProjector{n}\otimes\II),
\end{equation}
where $\numberOperator_{\mainMode{j}}$ is the photon-number operator on mode $\mainMode{j}$, $\mainProjector{n}$ acts on the main Fock space $\mainSpace$, and $\indicator[\mainNumberOp=n]$ denotes its spectral projector for the eigenvalue $n$. Note that the channel $\numberChannel$ acts on the joint system $\mainSystem$, and so only the \textit{total} photon number is virtually measured---measuring the two path occupations separately would destroy the interference within the single-photon subspace.

The key observation is that the nonselective measurement $\numberChannel$ can be inserted virtually before Bob's measurement without altering anything that Eve can observe. Indeed, the positive-operator-valued measure (POVM) elements of the main and satellite measurements commute with $\mainNumberOp$, because the phase shift and the exit beam splitter preserve the total photon number in $\mainSystem$ and the threshold detectors only distinguish photon numbers. Hence, $\numberChannel$ leaves the detection statistics unchanged and, conditioned on each detection record, so does the joint state of Eve's quantum memory and of all other unmeasured systems. Moreover, under \cref{asm:leak}, each leakage state is freshly prepared from the detection record and the receiver setting only, and thus the joint state of the detection records, the emitted leakage, and these unmeasured systems is also unmodified. We may therefore assume, without loss of generality, that Bob measures $\mainNumberOp$ and obtains an outcome $\mainNumber$. In particular, the single-photon subspace of the main modes is specified by the virtual outcome $\mainNumber=1$ and the qubit space $\qubitSpace:=\operatorname{span}\{|1,0\rangle_{\mainMode{1}\mainMode{2}},|0,1\rangle_{\mainMode{1}\mainMode{2}}\}$.
We denote this qubit by $\bobSystem$. Events with $\mainNumber\ne1$ are tagged, in the sense introduced above.

\subsection{Lower bound on the single-photon probability in the main modes}
\label{subsec:drv-prob}

Let $\fullHistory{i}$ denote the classical history containing the choices and measurement records assigned to all rounds up to and including round $i$ (we refer the reader to \cref{app:delayed} for a formal definition). This means that $\fullHistory{i-2}$ excludes all detection outcomes involving the current $i$-th pair, including the shared satellite record $\satRecord{i-1/2}$, which will allow us to analyze the odd and even rounds separately in \cref{app:delayed}. For any history of nonzero probability, and for fixed values of Alice's basis $w_{\aliceSystem}$, Bob's basis $w_{\bobSystem}$, and intensity $\meanPhoton$ in round $i$, we define the conditional probabilities of a main click, of a main click jointly with no satellite clicks, and of the target condition $\mainNumber=1$, respectively, as
\begin{equation}\label{eq:MTJ}
\begin{aligned}
 \conditionalMain &:=\Pr[\mainEvent^{(i)}\mid\fullHistory{i-2},w_{\aliceSystem},w_{\bobSystem},\meanPhoton],\\
 \conditionalJoint &:=\Pr[\mainEvent^{(i)},\satVacEvent^{(i)}\mid\fullHistory{i-2},w_{\aliceSystem},w_{\bobSystem},\meanPhoton],\\
 \conditionalTarget &:=\Pr[\mainNumber^{(i)}=1\mid\fullHistory{i-2},w_{\aliceSystem},w_{\bobSystem},\meanPhoton].
\end{aligned}
\end{equation}
For simplicity of notation, we omit the dependence of $\conditionalMain$, $\conditionalJoint$ and $\conditionalTarget$ on the history and on the setting. We stress, however, that the bound below holds conditionally on both, and does not require them to be statistically independent.

\begin{Thm}[Main single-photon lower bound]
\label{thm:single_bound}
Under \cref{asm:optical,asm:causal}, for every conditional input state $\density_{\earlyInputMode\lateInputMode}$ of the target pulse pair (see \cref{subsec:asm}), and for arbitrary states of the extra modes and of Eve's systems,
\begin{equation}\label{eq:single_bound}
 \conditionalTarget \geq\left[\frac{\conditionalJoint -\multiCoefficient \conditionalMain }{\satDarkFactor-\multiCoefficient}-\vacuumClick \right]_+,
\end{equation}
where $[x]_+:=\max\{0,x\}$ and
\begin{equation}\label{eq:q_explicit}
\begin{gathered}
 \satDarkFactor:=(1-\darkProb^{-})^4,
 \qquad
 \multiCoefficient:=\frac{\satDarkFactor \maxRouting ^2}{1-(1-\darkProb^{-})^2(1-\maxRouting )^2},
 \qquad 
 \maxRouting :=\max\{\entranceTrans_{+},1-\entranceTrans_{-}\},
 \qquad 
 \vacuumClick :=1-(1-\darkProb^{+})^2.
\end{gathered}
\end{equation}
%
\end{Thm}

The proof is provided in \cref{app:proof_lmm_qn}. In~\cref{eq:q_explicit}, $\vacuumClick$ is an upper bound on the probability that at least one detector clicks in the main window when the main modes are in vacuum, $\satDarkFactor$ is an upper bound on the probability that no dark count occurs at either detector in either of the two satellite windows, $\maxRouting$ is the largest probability with which an input photon may be routed to the main modes, and $\multiCoefficient$ is the coefficient that bounds the contribution of multiphoton inputs in the proof.
We remark that \cref{thm:single_bound} holds for any values of $\entranceTrans\in[\entranceTrans_-,\entranceTrans_+]$ and $\darkProb\in[\darkProb^-,\darkProb^+]$, and the users do not need to know the exact values of $\entranceTrans$ and $\darkProb$. Since the bound depends on these parameters only through the endpoints of their calibration intervals, it applies if $\entranceTrans$ varies between pulse pairs within its interval. Moreover, it also applies when $\darkProb$ varies between physical detection windows, provided that the independent, input-independent dark-count model of \cref{asm:optical} is retained---since the proof uses only the uniform bounds in \cref{eq:q_explicit}.
In practice, $\darkProb^-$ may be set to zero at a negligible cost, so that only an upper bound on the dark-count probability is actually required.
We stress that $\conditionalTarget$ refers to the photon number in the main modes rather than at the input of the receiver: a multiphoton input may leave exactly one photon in the main modes, while a single input photon may be routed entirely to a satellite mode.

We also note that a bound of a similar nature was derived in Ref.~\cite{Li2020FlagStateSquashing} from cross-click statistics, for the weight of the low-photon-number subspace at the input of a fully characterized receiver that processes each pulse pair in isolation. \Cref{thm:single_bound} instead bounds the probability that a specific pair of modes \textit{inside} the receiver---the main modes---contains exactly one photon, as required by the leakage analysis, and it holds for calibration intervals on $\entranceTrans$ and $\darkProb$, an uncharacterized exit beam splitter, and satellite windows shared with neighboring pairs whose pulses may be prepared adaptively by Eve.
\subsection{Virtual receiver and phase errors}
\label{subsec:virtual}

To accommodate information leakage from the receiver, we consider a virtual receiver that reproduces the actual leakage while delaying Bob's private bit readout, such that this readout can be replaced by a phase-error measurement. Specifically, we implement the leakage-isometry construction of Ref.~\cite{leakyreceiver2026} for the physical main-window patterns, and retain the systems associated with the satellite modes.

\subsubsection{Source replacement and target events}

For Alice's single-photon component, we introduce a qubit ancilla system $\aliceSystem$ and the purification
\begin{equation}\label{eq:alicevem}
|\replacementKetSymbol\rangle_{\aliceSystem\signalSystem}:=\frac{|0\rangle_{\aliceSystem}|1,\signalPhase=0\rangle_{\signalSystem}
                         +|1\rangle_{\aliceSystem}|1,\signalPhase=\pi\rangle_{\signalSystem}}{\sqrt2}.
\end{equation}
Note that the signal states are the $n=1$ states of \cref{eq:nphiS}. Measuring system $\aliceSystem$ in \cref{eq:alicevem} in the computational basis randomly prepares one of the two $Z$ signals. Measuring it instead in the orthonormal basis $\{|a_0\rangle_{\aliceSystem},|a_1\rangle_{\aliceSystem}\}$, with
\begin{equation}\label{eq:alice-complement}
 |a_0\rangle_{\aliceSystem}:=\frac{|0\rangle+i|1\rangle}{\sqrt2},\qquad
 |a_1\rangle_{\aliceSystem}:=\frac{|0\rangle-i|1\rangle}{\sqrt2},
\end{equation}
and obtaining the outcome $a\in\{0,1\}$ prepares, up to a global phase, the $X$ signal with bit value $a$. This measurement is complementary to Alice's key measurement.

\subsubsection{The three main patterns and their leakage}

Let $\hat{U}^{w}_{B\to D'}$ denote the unitary that maps the single-photon states of the main modes $m_1,m_2$ to those of the two output ports of the exit beam splitter, including the phase shift for basis $w$, and let $|j\rangle_{D'}$ denote the state with one photon in the output port leading to detector $D_j$. Since the ideal detector $D_j$ clicks precisely when the photon exits through its port, the projectors corresponding to the photon reaching detector $D_j$, expressed on the input space $\mathcal{H}_B$, are
\begin{equation}\label{eq:projectors}
    \hat{\Pi}^{j,w}_{B} := \big(\hat{U}^{w}_{B\to D'}\big)^{\dagger}|j\rangle_{D'}\langle j|\,\hat{U}^{w}_{B\to D'},
\end{equation}
with $\hat{\Pi}^{0,w}_{B}+\hat{\Pi}^{1,w}_{B}=\mathbb{I}_B$. Since all photon numbers are defined after the common loss, $N_m=1$ guarantees a click at the detector that receives the photon, and a double click occurs if the other detector produces a dark count. If $\hat{H}^{r,w}_{B}$ denotes the POVM element on $B$ associated with the pattern $r$, then we have that
\begin{equation}\label{eq:pattern-povm}
\begin{gathered}
    \hat{H}^{10,w}_{B}=(1-p_d)\hat{\Pi}^{0,w}_{B},\qquad \hat{H}^{01,w}_{B}=(1-p_d)\hat{\Pi}^{1,w}_{B},\qquad
    \hat{H}^{11,w}_{B}=p_d\,\mathbb{I}_B.
\end{gathered}
\end{equation}
Note that these elements sum to $\mathbb{I}_B$, and so no post-selection on a main click is needed for the target events: the physical detector loss is already accounted for in the probability of $N_m=1$. On the other hand, the POVM element for Bob's final recorded bit $b\in\{0,1\}$, after assigning a uniformly random bit to each double click, is given by
\begin{equation}\label{eq:bit-povm}
    \hat{G}^{b,w}_{B} := (1-p_d)\hat{\Pi}^{b,w}_{B}+\frac{p_d}{2}\,\mathbb{I}_B.
\end{equation}

Importantly, since $\hat{H}^{11,w}_{B}=p_d\mathbb{I}_B$, whether a round produces a single or a double click does not depend on the input qubit or on the basis. Indeed, $p_d$ is the actual dark-count probability in the current main window, which the users do not need to know, as the construction below holds for any $p_d\in[\darkProb^-,\darkProb^+]$. This branch decision---i.e., whether a round produces a single or a double click---can therefore be made before the final measurement, by drawing an independent classical random variable $\virtualBranch\in\{\text{single,double}\}$ with probabilities $(1-p_d,p_d)$, respectively. In the $\mathrm{single}$ branch, a projective outcome $j$ of $\{\hat{\Pi}^{j,Z}_B\}_j$ gives the single-click pattern $r=10$ for $j=0$ and $r=01$ for $j=1$. In the $\mathrm{double}$ branch, Bob records the pattern $r=11$ and obtains $b$ by substituting the state of the system $B$ by a maximally mixed state $\frac{1}{2}\,\mathbb{I}_B$ and measuring it. In both cases, the receiver then emits the state $\hat{\sigma}^{(r,Z)}_{L_m}$ associated with the recorded pattern. 

\subsubsection{Virtual receiver}

We now construct an equivalent virtual receiver in which the emission is generated \emph{before} the final readout. 
For this, in the $\mathrm{single}$ branch, by \cref{eq:Bobleaky} and the purification reduction in Appendix~D of Ref.~\cite{leakyreceiver2026}, we choose purifications $|f_0\rangle_F$ and $|f_1\rangle_F$ of $\hat{\sigma}^{(10,Z)}_{L_m}$ and $\hat{\sigma}^{(01,Z)}_{L_m}$, respectively, with real overlap $\langle f_0|f_1\rangle=\gamma$, where $F$ contains $L_m$ and the purifying auxiliaries; a fidelity lower bound is reduced to this exact overlap by means of an additional auxiliary system. 
With this, we define an isometry from $B$ to $BF$, which is applied before any measurement in the single branch, as
\begin{equation}\label{eq:isometry}
    \hat{W}^{Z}_{B\to BF} :=\hat{\Pi}^{0,Z}_{B}\otimes|f_0\rangle_F+\hat{\Pi}^{1,Z}_{B}\otimes|f_1\rangle_F, 
\end{equation}
which satisfies $(\hat{W}^{Z}_{B\to BF})^{\dagger}\hat{W}^{Z}_{B\to BF}=\mathbb{I}_B$. In the double branch, Bob emits the leakage state $\hat{\sigma}^{(11,Z)}_{L_m}$, and then substitutes the state of system $B$ by a maximally mixed state $\frac{1}{2}\mathbb{I}_B$. Finally, in both branches he measures $\{\hat{\Pi}^{b,Z}_B\}_b$ on $B$ to obtain $b$. 

Importantly, the virtual receiver above is at least as favorable to Eve as the actual one: the measurement statistics are identical, and the original state of the leakage system can be recovered by simply tracing out the purifying system in $F$ in the single branch.

\subsubsection{The phase-error operator}

The phase-error rate of the key rounds is defined through a virtual protocol in which Alice measures her target systems in the complementary basis instead of performing her key readouts, while Bob can perform any measurement he wishes. Importantly, in these rounds, either the leakage isometry $\hat{W}^{Z}_{B\to BF}$ or the state substitution acts on $B$ before Bob's readout, and thus the virtual measurement is applied to the modified system $B$. That is, if we trace out system $F$, the transformation is given by
\begin{equation}\label{eq:branch-channel}
    \hat{\rho}\to (1-p_d)\mathcal{E}^{\gamma}_{B}(\hat{\rho})+\frac{p_d}{2}\mathbb{I}_B \operatorname{Tr}_B[\hat{\rho}]
    = (1-p_d)\mathcal{E}^{\gamma}_{B}(\hat{\rho})+\frac{p_d}{2}\mathbb{I}_B \operatorname{Tr}_B[\mathcal{E}^{\gamma}_{B}(\hat{\rho})],
\end{equation}
where $\hat{\rho}$ is any input state on system $B$ and
\begin{equation}\label{eq:dephasing}
    \mathcal{E}^{\gamma}_{B}(\hat{\rho}) := \operatorname{Tr}_F\Big[\hat{W}^{Z}_{B\to BF}\,\hat{\rho}\,\big(\hat{W}^{Z}_{B\to BF}\big)^{\dagger}\Big]
    = \sum_{j=0}^{1}\hat{\Pi}^{j,Z}_{B}\hat{\rho}\,\hat{\Pi}^{j,Z}_{B}
    + \gamma\sum_{j\neq k}\hat{\Pi}^{j,Z}_{B}\hat{\rho}\,\hat{\Pi}^{k,Z}_{B},
\end{equation}
is a self-adjoint, trace-preserving dephasing channel, and $\gamma$ is introduced in Eq.~\eqref{eq:Bobleaky}.
The equality in \cref{eq:branch-channel} shows that we can equivalently regard the effective dephasing as acting on the input state in both branches, followed by the branch-dependent readout, whose POVM elements are $\hat{G}^{b,Z}_{B}$. We remark that tracing out $F$ in Eq.~(\ref{eq:dephasing}) is used only to calculate the error probabilities; $F$ remains available to Eve in the security proof. 

Now let us define the $X$-basis single-photon bit-error operator as
\begin{equation}\label{eq:error-operators}
    \hat{E}^{X}_{AB} := |a_0\rangle_A\langle a_0|\otimes\hat{G}^{1,X}_{B}+|a_1\rangle_A\langle a_1|\otimes\hat{G}^{0,X}_{B},
\end{equation}
where $\hat{G}^{b,w}_{B}$ is defined in Eq.~(\ref{eq:bit-povm}). The operator $\hat{E}^{X}_{AB}$ corresponds to a bit error in a test round, in which Alice obtains the outcome $a$ in the basis of \cref{eq:alice-complement} and Bob records the bit $1-a$ in the $X$ basis. Then, if we replace Bob's readout after the dephasing by the test measurement in~\cref{eq:error-operators}, the effective phase-error operator is given by
\begin{equation}\label{eq:phase-error-operators}
    \hat{E}^{\rm ph}_{AB} := (\mathrm{id}_A\otimes\mathcal{E}^{\gamma}_{B})(\hat{E}^{X}_{AB}),
\end{equation}
where $\mathrm{id}_A$ is the identity channel on $A$. The difference between the two error operators is therefore due to the dephasing $\mathcal{E}^{\gamma}_{B}$ alone, which we bound next.

To bound the difference between the two error operators, we write $\hat{E}^{\rm ph}_{AB}-\hat{E}^{X}_{AB}=\sum_{a=0}^{1}|a_a\rangle_A\langle a_a|\otimes\hat{\Delta}_a$, with $\hat{\Delta}_a:=(\mathcal{E}^{\gamma}_{B}-\mathrm{id}_B)(\hat{G}^{1-a,X}_{B})$. Since the projectors $|a_0\rangle\langle a_0|$ and $|a_1\rangle\langle a_1|$ are orthogonal, this operator is block diagonal, and thus its operator norm is $\max_a\|\hat{\Delta}_a\|_\infty$, where $\|\cdot\|_\infty$ denotes the operator norm. To evaluate $\hat{\Delta}_a$, note that, in the orthonormal basis defined by $\{\hat{\Pi}^{0,Z}_B,\hat{\Pi}^{1,Z}_B\}$, the dephasing channel $\mathcal{E}^{\gamma}_{B}$ leaves the diagonal entries of an operator unchanged and multiplies its off-diagonal entries by $\gamma$. Hence, the term $p_d\mathbb{I}_B/2$ of $\hat{G}^{1-a,X}_{B}$ in Eq.~(\ref{eq:bit-povm}) does not contribute to $\hat{\Delta}_a$, which equals $(1-p_d)(\gamma-1)$ times the off-diagonal part of $\hat{\Pi}^{1-a,X}_B$. Finally, since the off-diagonal part of a qubit projector has operator norm at most $1/2$, we obtain
\begin{equation}\label{eq:delta}
    \|\hat{E}^{\rm ph}_{AB}-\hat{E}^{X}_{AB}\|_\infty \leq \frac{(1-p_d)(1-\gamma)}{2} =: \delta.
\end{equation}
Here, $p_d$ is the dark-count probability of the main window of the round under consideration, so $\delta$ may differ between rounds. Since $p_d\geq\darkProb^-$, however, $\delta\leq\overline{\phasePenalty}:=(1-\darkProb^-)(1-\leakOverlap)/2$ in every round, and it is this computable uniform bound that enters the key rate below.
Note that Bob's two bases need not be mutually unbiased. In particular, arbitrary exit beam-splitter imbalance and a fixed main phase offset are compatible with this bound, as the complementarity required for the uncertainty relation is supplied by Alice's two measurements.

\subsubsection{Phase-error rate of the target events}
We now use the error operators above to relate the phase-error rate of the key rounds to the bit-error rate of the test rounds. For this, we need the joint state of Alice's qubit and Bob's main modes in the target events to be the same in both basis experiments. Importantly, the target condition $\mainNumber=1$ does not depend on Bob's basis choice, because the total photon number in $\mainSystem$ is defined before the long-arm phase shift and the exit beam splitter act. Moreover, Alice's basis can be chosen virtually after transmission, because her bit-averaged single-photon state is basis independent, and Bob's basis is independent of the protected input by \cref{asm:causal}. Specifically, let us fix the history $\fullHistory{i-2}$ and denote by $\targetState^{(i)}$ the joint state of $\aliceSystem$ and $\bobSystem$ in round $i$ conditioned on the target event $(1_{\aliceSystem},1_m)$, weighted by its conditional probability, such that $\operatorname{Tr}\targetState^{(i)}=\Pr(\alicePhotonNumber=1,\mainNumber=1\mid\fullHistory{i-2})$, where $\alicePhotonNumber$ is Alice's emitted photon number. This state is evaluated after Bob's virtual measurement of $\mainNumberOp$, but before the basis measurements and the leakage isometry of round $i$, and with the unrevealed intensity label averaged over. By the arguments above, $\targetState^{(i)}$ is the same for the key and the test basis choices. In addition, since the key rounds are retained regardless of the satellite outcomes (see \cref{step:sifting} of the protocol), and since the satellite measurements and leakage act on systems other than $\aliceSystem\bobSystem$, summing over the satellite outcomes leaves $\targetState^{(i)}$ unchanged, while all satellite leakage remains available to Eve. We remark that neither Alice nor Bob measures the target label in the actual protocol; it is only used in the security proof.

Let $\conditionalPhaseError{i}$ and $\conditionalTestError{i}$ denote the conditional probabilities, given $\fullHistory{i-2}$, of a phase error in a basis-matched key round and of a bit error in a basis-matched test round, respectively, both restricted to the target events. By \cref{eq:error-operators,eq:phase-error-operators,eq:delta} and the positivity of $\targetState^{(i)}$, we have that
\begin{equation}\label{eq:local-phase}
\begin{aligned}
 \conditionalPhaseError{i}
 &:=\basisProbability{Z}^2\operatorname{Tr}[\phaseErrorEffect\targetState^{(i)}]\\
 &\leq \basisProbability{Z}^2\operatorname{Tr}[\testErrorEffect\targetState^{(i)}]
       +\phasePenalty \basisProbability{Z}^2\operatorname{Tr}\targetState^{(i)}\\
 &=\frac{\basisProbability{Z}^2}{\basisProbability{X}^2}\conditionalTestError{i}
       +\phasePenalty \basisProbability{Z}^2\operatorname{Tr}\targetState^{(i)},
\end{aligned}
\end{equation}
with $\conditionalTestError{i}:=\basisProbability{X}^2\operatorname{Tr}[\testErrorEffect\targetState^{(i)}]$, where the inequality follows from $\operatorname{Tr}[(\phaseErrorEffect-\testErrorEffect)\targetState^{(i)}]\leq\|\phaseErrorEffect-\testErrorEffect\|_\infty\operatorname{Tr}\targetState^{(i)}$, and the factors $\basisProbability{Z}^2$ and $\basisProbability{X}^2$ account for the basis-matching probabilities.

To convert these conditional probabilities into counts, we define $\targetGain^{w}$ as the number of basis-matched target events in basis $w$, divided by $\totalPairs$. Likewise, $\testErrorYield^{X}$ and $\phaseErrorYield^{Z}$ are the corresponding test-bit-error and virtual phase-error counts divided by $\totalPairs$, and $\targetTestRate^{X}:=\testErrorYield^{X}/\targetGain^{X}$ and $\targetPhaseRate:=\phaseErrorYield^{Z}/\targetGain^{Z}$ are the corresponding error rates within the target events. By the basis independence of $\targetState^{(i)}$ established above, the conditional probability of a basis-matched target event in basis $w$ is $\basisProbability{w}^2\operatorname{Tr}\targetState^{(i)}$. The concentration argument in \cref{app:delayed} thus shows that, asymptotically, $\targetGain^{w}$ equals $\sum_i\basisProbability{w}^2\operatorname{Tr}\targetState^{(i)}/\totalPairs$, and similarly for the error counts. Summing \cref{eq:local-phase} over the rounds, and bounding $\phasePenalty$ by $\overline{\phasePenalty}$ in each of them, we therefore obtain
\begin{align}
 \frac{\targetGain^{Z}}{\basisProbability{Z}^2}&=\frac{\targetGain^{X}}{\basisProbability{X}^2},\label{eq:basis-ratio}\\
 \phaseErrorYield^{Z}&\leq\frac{\basisProbability{Z}^2}{\basisProbability{X}^2}\testErrorYield^{X}
                             +\overline{\phasePenalty} \targetGain^{Z},
                             \label{eq:phase-yield}\\
 \targetPhaseRate&\leq \targetTestRate^{X}+\overline{\phasePenalty}.\label{eq:phase-rate}
\end{align}
That is, the leakage contribution to the phase-error count scales with the number of target events, and the resulting additive correction $\overline{\phasePenalty}$ to the phase-error rate is independent of the channel transmittance.

\subsection{Target-yield and phase-error estimation}
\label{subsec:estimation}

We now express the virtual target quantities in terms of the observed data. Let $n_{w,\meanPhoton}$ be the number of main clicks with matching basis $w$ and intensity $\meanPhoton$, let $\widetilde n_{w,\meanPhoton}$ be the number of those for which both neighboring satellite records are $00$, and let $m_{X,\meanPhoton}$ be the number of bit errors among the test-round bits with intensity $\meanPhoton$ (see \cref{step:sifting,step:estimation} of the protocol). We define the observed gains and error yields, normalized by $\totalPairs$, as
\begin{equation}\label{eq:observed-gains}
 \observedGain{w}{\meanPhoton}:=\frac{n_{w,\meanPhoton}}{\totalPairs },\qquad
 \jointGain_{w,\meanPhoton}:=\frac{\widetilde n_{w,\meanPhoton}}{\totalPairs },\qquad
 \totalGain{w}:=\sum_\meanPhoton \observedGain{w}{\meanPhoton},\qquad
 \observedErrors^{X,\meanPhoton}:=\frac{m_{X,\meanPhoton}}{\totalPairs}.
\end{equation}
Moreover, let $\mainGain^{w,\meanPhoton}$ denote the number of $w$-basis-matched rounds with intensity $\meanPhoton$ in which Bob's main modes contain exactly one photon, divided by $\totalPairs$. This quantity is not observed in the actual protocol, but applying \cref{thm:single_bound} to each round, summing over the rounds, and using the concentration argument in \cref{app:delayed}, it can be bounded as
\begin{equation}\label{eq:Q1B_lower}
 \mainLower^{w,\meanPhoton}:=\left[\frac{\jointGain_{w,\meanPhoton}-\multiCoefficient \observedGain{w}{\meanPhoton}}{\satDarkFactor-\multiCoefficient}
                         -\basisProbability{w}^2\intensityProbability \vacuumClick \right]_+
 \leq \mainGain^{w,\meanPhoton}\leq \observedGain{w}{\meanPhoton},
\end{equation}
where $\satDarkFactor$, $\multiCoefficient$, and $\vacuumClick$ are defined in \cref{eq:q_explicit}. The vacuum correction $\vacuumClick$ is multiplied by $\basisProbability{w}^2\intensityProbability$ because the gains are normalized by the total number of transmitted pairs $\totalPairs$, of which a fraction $\basisProbability{w}^2\intensityProbability$ asymptotically has the specified basis and intensity choices, irrespective of their detection outcomes.

Next, to isolate the single-photon contribution of Alice's source, we use the decoy-state method~\cite{Lo2005DecoyState}. For this, we define
\begin{equation}\label{eq:decoy-probs}
 \emissionProbability{k}:=\sum_{\meanPhoton\in\intensitySet}\intensityProbability \photonProbability{k},\qquad
 \posteriorIntensity{k}:=\frac{\intensityProbability \photonProbability{k}}{\emissionProbability{k}},
\end{equation}
where $\emissionProbability{k}$ is the total probability that Alice emits $k$ photons, and $\photonProbability{k}$ is the Poisson probability in \cref{eq:poisson}. Since, at fixed $k$, the emitted state does not depend on the intensity label (see \cref{subsec:source}), this label can be regarded as being sampled virtually after the optical processing with probability $\posteriorIntensity{k}$. Writing $\photonTargetGain{k}^{w}$ for the number of basis-matched rounds in which Alice emits $k$ photons and Bob's main modes contain one photon, divided by $\totalPairs$, this yields
\begin{equation}\label{eq:virtual_expansion}
 \mainGain^{w,\meanPhoton}=\sum_{k\geq0}\posteriorIntensity{k}\photonTargetGain{k}^{w}.
\end{equation}
Thus, combining \cref{eq:virtual_expansion} with the bounds in \cref{eq:Q1B_lower}, the standard three-intensity decoy-state analysis~\cite{charles2013} gives $\targetGain^{w}\geq\targetLower^{w}$, where
\begin{equation}\label{eq:Q1_lower_bound}
 \targetLower^{w}
 :=\Bigg[\frac{\meanPhoton_1\emissionProbability{1}}{\meanPhoton_1(\meanPhoton_2-\meanPhoton_3)-(\meanPhoton_2^2-\meanPhoton_3^2)}
 \Bigg(\frac{e^{\meanPhoton_2}\mainLower^{w,\meanPhoton_2}}{\intensityChoiceProbability{2}}
       -\frac{e^{\meanPhoton_3}\observedGain{w}{\meanPhoton_3}}{\intensityChoiceProbability{3}}
       -\frac{\meanPhoton_2^2-\meanPhoton_3^2}{\meanPhoton_1^2}
        \frac{e^{\meanPhoton_1}\observedGain{w}{\meanPhoton_1}}{\intensityChoiceProbability{1}}\Bigg)\Bigg]_+.
\end{equation}
Note that, since the single-photon gains $\mainGain^{w,\meanPhoton}$ are not directly observed, we have used their lower bounds from \cref{eq:Q1B_lower} in the positive term of the decoy expression, and their upper bounds $\observedGain{w}{\meanPhoton}$ in the negative terms.

Similarly, to upper bound the test-error yield $\testErrorYield^{X}$, let $\alicePhotonErrors{k}^{X}$ denote the number of test-round bit errors in which Alice emits $k$ photons, divided by $\totalPairs$, with no restriction on Bob's photon number. The same sampling argument gives
\begin{equation}\label{eq:error_virtual_expansion}
 \observedErrors^{X,\meanPhoton}=\sum_{k\geq0}\posteriorIntensity{k}\alicePhotonErrors{k}^{X},\qquad
 \testErrorYield^{X}\leq \alicePhotonErrors{1}^{X}.
\end{equation}
Taking the difference between the $\meanPhoton_2$ and $\meanPhoton_3$ expressions cancels the vacuum contribution, and dropping the nonnegative multiphoton terms gives
\begin{equation}\label{eq:E11_upper_bound}
 \testErrorYield^{X}\leq
 \testErrorUpper:=\frac{\emissionProbability{1}}{\meanPhoton_2-\meanPhoton_3}
 \left(\frac{e^{\meanPhoton_2}\observedErrors^{X,\meanPhoton_2}}{\intensityChoiceProbability{2}}
       -\frac{e^{\meanPhoton_3}\observedErrors^{X,\meanPhoton_3}}{\intensityChoiceProbability{3}}\right).
\end{equation}
This upper bound is nonnegative for exact asymptotic data.

Finally, combining \cref{eq:phase-rate} with the bounds above, the phase-error rate of the target events satisfies $\targetPhaseRate\leq\testErrorUpper/\targetLower^{X}+\overline{\phasePenalty}$ whenever $\targetLower^{X}>0$, with $\overline{\phasePenalty}$ the leakage penalty introduced below \cref{eq:delta}.
For the key-rate formula below, we define
\begin{equation}\label{eq:eph}
 \entropyPhase:=\min\left\{\frac12,
 \frac{\testErrorUpper}{\targetLower^{X}}+\overline{\phasePenalty}\right\}.
\end{equation}
Since $\overline{\phasePenalty}\leq(1-\leakOverlap)/2$, the simpler penalty $(1-\leakOverlap)/2$ may replace $\overline{\phasePenalty}$ at the cost of a slightly looser bound.

\subsection{Asymptotic key rate}
\label{sec:derive_rate}

We are now ready to derive the key rate. As explained at the beginning of this section, we conservatively give Eve the target labels and the bit values of all nontarget rounds, in addition to all receiver emissions, the auxiliary virtual systems $\purificationSystem$ and $\virtualBranch$, and the public transcript. Applying the entropic uncertainty relation to Alice's complementary measurements and the leftover hash lemma~\cite{tomamichel2011,tomamichel2017,tomamichel2011hashing}, the asymptotic secret-key rate per transmitted pair satisfies $R\geq\keyRate$, with
\begin{equation}\label{eq:key_rate_final}
 \keyRate:=\left[\targetLower^{Z}(1-\binaryEntropy (\entropyPhase))-\ecDisclosure\right]_+,
\end{equation}
where $\binaryEntropy(x):=-x\log_2x-(1-x)\log_2(1-x)$ is the binary entropy function and $\ecDisclosure$ is the error-correction disclosure per transmitted pair, for which we use the common model $\ecDisclosure=\ecInefficiency \totalGain{Z}\binaryEntropy (\keyBitError)$, with $\ecInefficiency\geq1$ the reconciliation inefficiency and $\keyBitError$ the observed key-basis bit-error rate. We set $\keyRate=0$ whenever $\targetLower^{Z}$ or $\targetLower^{X}$ vanishes. Authentication and verification require a number of bits sublinear in $\totalPairs$ with vanishing failure probability, and therefore do not affect the asymptotic rate.

\section{Numerical results}
\label{sec:sim}

\subsection{Tightness of the single-photon bound}

For simplicitly, in this section, we assume exact calibration, i.e., $\entranceTrans_\pm=\entranceTrans$ and $\darkProb^\pm=\darkProb$, so that the coefficients in \cref{eq:q_explicit} take their exact values and $\overline{\phasePenalty}=\phasePenalty$. We first evaluate the tightness of the bound in \cref{thm:single_bound} using detection statistics generated by a pure-loss channel. For this, we consider a continuous train of independently phase-randomized pairs, each with mean photon number $\meanPhoton$ equally divided between its early and late pulses, and the shared-window geometry of \cref{eq:shared-record}. The effective transmittance at a distance $d$ is given by $t(d):=\detEfficiency10^{-0.2d/10}$,
where $d$ is in kilometers, the attenuation coefficient is $0.2\,\mathrm{dB/km}$, and $\detEfficiency$ is the detector efficiency. After the common loss, the main modes contain a total mean photon number $\receivedMean/2$, with $\receivedMean:=t(d)\meanPhoton$, which is the same mean photon number each satellite window receives, if we include the contributions of the neighboring pairs. Since both output ports are monitored, the no-click probability of the satellite windows does not depend on the interference phase. Hence,
\begin{equation}\label{eq:simulation-estimator}
 \conditionalMain =1-(1-\darkProb)^2e^{-\receivedMean/2},\qquad
 \conditionalJoint =\satDarkFactor e^{-\receivedMean}\conditionalMain ,\qquad
 \conditionalTargetExact:=\frac{\receivedMean}{2}e^{-\receivedMean/2},
\end{equation}
where $\conditionalTargetExact$ is the exact probability that the main modes contain one photon. These expressions follow from the splitting of coherent states at the beam splitters and from the independent phase randomization of each pair.

\begin{figure}[tb]
  \centering
  \includegraphics[scale=1.0]{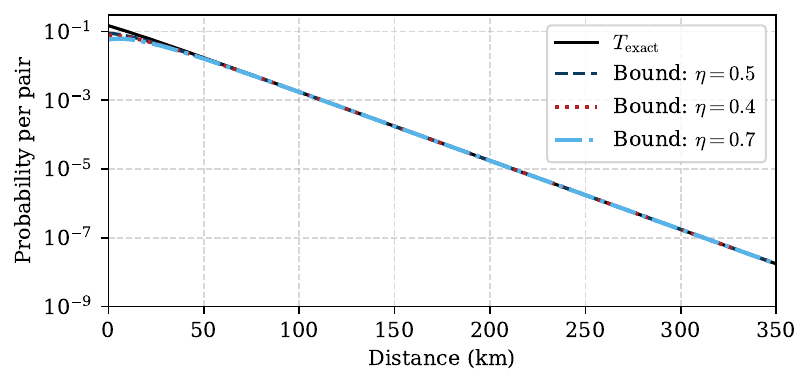}
  \caption{Main-mode single-photon probability $\conditionalTargetExact$ from \cref{eq:simulation-estimator} and its lower bound from \cref{eq:single_bound}, with $\meanPhoton=0.5$, $\detEfficiency=0.7$, and $\darkProb=10^{-8}$. Both detector ports are monitored; the estimator does not depend on $\exitTrans$.}
  \label{fig:est_single_photon}
\end{figure}

\begin{figure}[tb]
  \centering
  \includegraphics[scale=1.0]{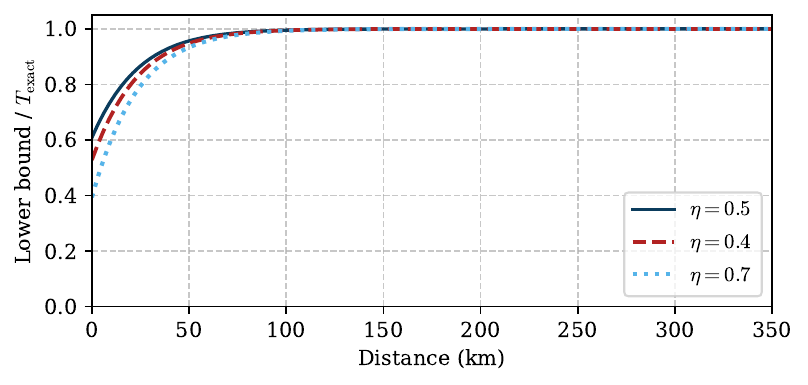}
  \caption{Ratio of the lower bound in \cref{eq:single_bound} to the exact main-mode single-photon probability $\conditionalTargetExact$ in \cref{eq:simulation-estimator}, for the parameters of \cref{fig:est_single_photon}. The curves approach the near-unit limits in \cref{eq:ratio-limit} as the channel loss increases.}
  \label{fig:est_ratio}
\end{figure}

\Cref{fig:est_single_photon,fig:est_ratio} show the lower bound in \cref{eq:single_bound} together with $\conditionalTargetExact$, and their ratio, for $\meanPhoton=0.5$, $\detEfficiency=0.7$, $\darkProb=10^{-8}$ and $\entranceTrans\in\{0.4,0.5,0.7\}$. Moreover, expanding \cref{eq:simulation-estimator} around $\receivedMean=0$ for fixed $\entranceTrans$ and $\darkProb$ gives
\begin{equation}\label{eq:ratio-limit}
 \lim_{\receivedMean\to0}\frac{[\,(\conditionalJoint -\multiCoefficient \conditionalMain )/(\satDarkFactor-\multiCoefficient)-\vacuumClick \,]_+}{\conditionalTargetExact}
 =\left[1-\vacuumClick -\frac{2\satDarkFactor \vacuumClick }{\satDarkFactor-\multiCoefficient}\right]_+,
\end{equation}
which equals one for $\darkProb=0$ and is close to one for the parameters above. That is, the bound in \cref{thm:single_bound} becomes essentially tight in the high-loss regime, as shown in \cref{fig:est_ratio}.

\subsection{Key-rate comparison}
\label{subsec:key-numerics}

We now evaluate the key-rate bound in \cref{eq:key_rate_final}. For this, we simulate the observed statistics with the channel model of \cref{sec:sim_setup}, which includes the interference visibility of the imbalanced interferometer and the intensities of the neighboring pairs, drawn independently from the same distribution as the current pair. We use $\darkProb=10^{-8}$, $\detEfficiency=0.7$, $\basisProbability{Z}=0.8$, $(\intensityChoiceProbability{1},\intensityChoiceProbability{2},\intensityChoiceProbability{3})=(0.8,0.1,0.1)$, $\meanPhoton_3=0$, and $\ecInefficiency=1$, and, at each distance, we optimize $\meanPhoton_1$ and $\meanPhoton_2$ over the grid specified in \cref{sec:sim_setup}. We define the cutoff distance as the largest distance at which the optimized key-rate bound is positive.

\begin{figure}[tb]
  \centering
  \includegraphics[scale=1.0]{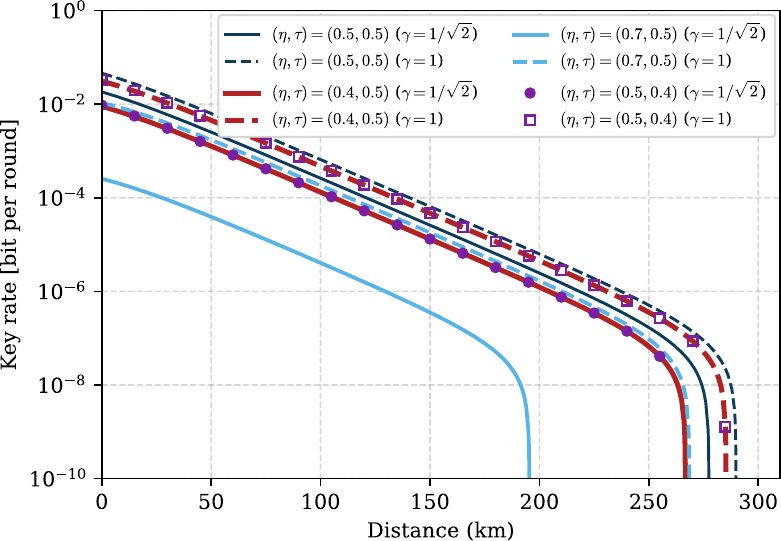}
  \caption{Asymptotic key-rate lower bounds from \cref{eq:key_rate_final} for the parameters in \cref{subsec:key-numerics} and the intensity grid in \cref{sec:sim_setup}. Solid (dashed) curves and filled circles (open squares) use $\leakOverlap=1/\sqrt2$ ($\leakOverlap=1$). Only positive rates are shown. The $(0.4,0.5)$ and $(0.5,0.4)$ curves nearly coincide in this model.}
  \label{fig:key-rate}
\end{figure}

\Cref{fig:key-rate} shows the results for $\leakOverlap=1/\sqrt{2}$ and $\leakOverlap=1$, and for several values of $(\entranceTrans,\exitTrans)$. We recall that $\leakOverlap=1$ corresponds to identical leakage states for the two single-click patterns in the key basis, while in both cases the satellite and double-click leakage specified in \cref{asm:leak} is given to Eve in full. We see that the beam-splitter imbalance reduces the key rate in two ways: through the optical error caused by the reduced visibility, and, in the case of the entrance imbalance and mainly at short distances, through the single-photon estimator, as $\entranceTrans$ enters the coefficients in \cref{eq:q_explicit}. Decreasing $\leakOverlap$, on the other hand, increases the leakage penalty $\overline{\phasePenalty}$ in the phase-error bound.

Importantly, the two effects compound. For the balanced receiver, $(\entranceTrans,\exitTrans)=(0.5,0.5)$, decreasing $\leakOverlap$ from $1$ to $1/\sqrt{2}$ reduces the cutoff distance from $289.88$ to $277.49\,\mathrm{km}$, i.e., by about $12\,\mathrm{km}$, whereas for $(\entranceTrans,\exitTrans)=(0.7,0.5)$ the same change reduces it from $268.53$ to $195.44\,\mathrm{km}$, i.e., by about $73\,\mathrm{km}$. The reason is that the optical error increases both the error-correction and the privacy-amplification costs, which leaves less margin for the additional phase-error penalty due to leakage.

We also note that the $(0.4,0.5)$ and $(0.5,0.4)$ curves nearly coincide. The reason is that the main-window visibility is symmetric under $\entranceTrans\leftrightarrow\exitTrans$ (see \cref{sec:sim_setup}), while the additional dependence on $\entranceTrans$ through the coefficient $\multiCoefficient$ of the single-photon estimator vanishes to first order in the received mean photon number when dark counts are neglected.

\section{Conclusions}

In summary, in this work we have derived an analytical key-rate lower bound for phase-encoded decoy-state BB84 with pattern-dependent receiver leakage. The key ingredient of the proof is an internal virtual measurement of the total photon number in the two optical modes that interfere in the main detection window, which identifies Bob's single-photon qubit contribution while preserving the joint state of the detection records and the emitted leakage. By bounding this contribution from the click statistics in the main and satellite windows, we have been able to apply a phase-error analysis that accommodates the information leaked about the measurement outcomes. The resulting leakage-induced correction to the phase-error rate is independent of channel loss, and our security proof accommodates imbalanced beam splitters, with an entrance transmittance known only within a calibration interval and a completely uncharacterized exit transmittance. 

Our numerical results show that, in the strong receiver-leakage regime considered here, beam-splitter imbalance can 
substantially amplify the degradation in the key rate and maximum transmission distance. 
Specifically, decreasing $\gamma$ from 1 to $1/\sqrt{2}$ reduces the maximum transmission distance by approximately 12 km for the balanced receiver, 
compared with approximately 73 km for $(\eta, \tau) = (0.7, 0.5)$.

As for directions for future research, these include extending the analysis to detection-efficiency mismatch and to receivers with memory, and obtaining a complete finite-key bound.

\section{Acknowledgements}

We thank Shun Kawakami, Guillermo Currás-Lorenzo, and Margarida Pereira for helpful discussions. A.M. was partially supported by JSPS KAKENHI Grant No. JP24K16977. K.T. acknowledges support from JSPS KAKENHI Grant Numbers 23K25793 and 23H01096. A.N. acknowledges financial support from Xunta de Galicia through a Postdoctoral Fellowship (No. ED481B-2025/113).

ChatGPT 6 Astra and Claude Opus 5.5/Fable 5.1 were used to assist with the refinement of some mathematical derivations and the preparation of the manuscript. All AI-assisted content was critically reviewed and verified by the authors, who take full responsibility for the entire manuscript.

\appendix

\section{Proof of Theorem \ref{thm:single_bound}}
\label[appendix]{app:proof_lmm_qn}
The proof has two steps. First, we express the three events in \cref{eq:MTJ}---a main click, a main click with no satellite clicks, and the target condition $\mainNumber=1$---as operators on the same target-input Fock space. Then, we establish an operator inequality between them, whose expectation value gives the conditional probability bound. To remove the dependence on the extra modes, we bound their vacuum projector by the identity.

\begin{proof}
On the full main Fock space $\mainSpace$, the POVM element corresponding to a main click is
\begin{equation}\label{eq:main-effect}
 \mainClickEffect:=\II_{\mainSystem}-(1-\darkProb)^2\mainProjector{0},
\end{equation}
where $\darkProb$ is the actual dark-count probability in the main window (see~\cref{eq:main-number} for the definition of the operators). Since both output ports of the satellite windows are monitored, destructive interference at one port cannot produce no clicks at both ports if any photon is present. Hence, pulling the no-click event back through the lossless exit beam splitters, the POVM element corresponding to no clicks in the two satellite windows is $\satDarkFactor^{\prime} \vacuumProjector{\satMode{1}\satMode{2}}\otimes \vacuumProjector{\extraMode{1}\extraMode{2}}$, where $\satDarkFactor'$ is the probability that no dark count occurs at either detector in either of the two satellite windows---a product of four factors $1-\darkProb$, which thus satisfies $\satDarkFactor'\leq\satDarkFactor$ (see~\cref{eq:q_explicit})---and $\vacuumProjector{\satMode{1}\satMode{2}}$ and $\vacuumProjector{\extraMode{1}\extraMode{2}}$ are the joint-vacuum projectors on the satellite and extra modes, respectively. Since $\vacuumProjector{\extraMode{1}\extraMode{2}}\leq\II_{\extraMode{1}\extraMode{2}}$, with $\II_{\extraMode{1}\extraMode{2}}$ the identity on $\extraMode{1}\extraMode{2}$, and $\satDarkFactor'\leq\satDarkFactor$, we have that
\begin{equation}\label{eq:extra-domination}
 \mainClickEffect\otimes \satDarkFactor^{\prime} \vacuumProjector{\satMode{1}\satMode{2}}\otimes \vacuumProjector{\extraMode{1}\extraMode{2}}
 \leq \mainClickEffect\otimes \satDarkFactor \vacuumProjector{\satMode{1}\satMode{2}}\otimes\II_{\extraMode{1}\extraMode{2}},
\end{equation}
which removes the dependence on the extra modes and on the actual satellite dark-count probabilities. Pulling the remaining operators back through the entrance isometry, we define the following three operators on the target-input Fock space $\inputSpace$:
\begin{equation}\label{eq:input-operators}
\begin{aligned}
 \inputClickEffect&:=\left(\entranceIsometry\right)^\dagger\left(\mainClickEffect\otimes\II_{\satelliteSystem}\right)\entranceIsometry,\\
 \inputTargetEffect&:=\left(\entranceIsometry\right)^\dagger\left(\mainProjector{1}\otimes\II_{\satelliteSystem}\right)\entranceIsometry,\\
 \inputJointEffect&:=\satDarkFactor\left(\entranceIsometry\right)^\dagger\left(\mainClickEffect\otimes\vacuumProjector{\satMode{1}\satMode{2}}\right)\entranceIsometry.
\end{aligned}
\end{equation}
For any joint state of the target and extra modes with target marginal $\density_{\earlyInputMode\lateInputMode}$, we then have that $\conditionalMain =\operatorname{Tr}[\inputClickEffect\density_{\earlyInputMode\lateInputMode}]$, $\conditionalTarget =\operatorname{Tr}[\inputTargetEffect\density_{\earlyInputMode\lateInputMode}]$, and, by \cref{eq:extra-domination}, $\conditionalJoint \leq\operatorname{Tr}[\inputJointEffect\density_{\earlyInputMode\lateInputMode}]$. Moreover, the three operators in \cref{eq:input-operators} are diagonal in the Fock basis $\{|n_e,n_l\rangle\}$ of $\inputSpace$, as the operators in \cref{eq:main-effect,eq:extra-domination} are diagonal in the Fock basis of the output modes and the entrance beam splitter conserves the photon number of each input mode separately, i.e., $n_e=n_{\mainMode{1}}+n_{\satMode{1}}$ and $n_l=n_{\mainMode{2}}+n_{\satMode{2}}$. Thus, it suffices to compare their diagonal entries. Denoting these by $d$, $t$ and $f$, respectively, for the occupation numbers $n_e,n_l$, a direct calculation gives
\begin{align}
 d&=1-(1-\darkProb)^2\entranceTrans^{n_e}(1-\entranceTrans)^{n_l},\label{eq:diag-d}\\
 t&=\entranceTrans^{n_e}(1-\entranceTrans)^{n_l}
       \left(n_e\frac{1-\entranceTrans}{\entranceTrans}+n_l\frac{\entranceTrans}{1-\entranceTrans}\right),\label{eq:diag-t}\\
 f&=\begin{cases}
 \satDarkFactor [1-(1-\darkProb)^2],&n_e+n_l=0,\\
 \satDarkFactor(1-\entranceTrans)^{n_e}\entranceTrans^{n_l},&n_e+n_l\geq1,
 \end{cases}\label{eq:diag-f}
\end{align}
where we have used that a photon in mode $\earlyInputMode$ ($\lateInputMode$) is routed to the main modes with probability $1-\entranceTrans$ ($\entranceTrans$), and that no satellite clicks require all photons to be routed to the main modes. We now consider three cases: (i) for a vacuum input, $t=0$ and $f-\multiCoefficient d=[1-(1-\darkProb)^2](\satDarkFactor-\multiCoefficient)\leq\vacuumClick(\satDarkFactor-\multiCoefficient)$, since $\darkProb\leq\darkProb^+$; (ii) for one input photon, $d=t+[1-(1-\darkProb)^2](1-t)$ and $f=\satDarkFactor t$, so that $f-\multiCoefficient d\leq(\satDarkFactor-\multiCoefficient)t$; finally, (iii) for $n:=n_e+n_l\geq2$, denoting by $x_k\in\{\entranceTrans,1-\entranceTrans\}$ the probability that the $k$-th photon is routed to the main modes, we have that
\begin{equation}\label{eq:multi-ratio}
 \frac{f}{d}=\frac{\satDarkFactor\prod_{k=1}^n x_k}{1-(1-\darkProb)^2\prod_{k=1}^n(1-x_k)}
 \leq\frac{\satDarkFactor \maxRouting ^n}{1-(1-\darkProb^{-})^2(1-\maxRouting )^n}\leq \multiCoefficient,
\end{equation}
where $\maxRouting$ is defined in~\cref{eq:q_explicit}, and upper bounds every $x_k$ because $\entranceTrans\in[\entranceTrans_-,\entranceTrans_+]$. In~\cref{eq:multi-ratio}, the first inequality holds because the ratio is nondecreasing in each $x_k$ and in $a:=(1-\darkProb)^2\leq(1-\darkProb^-)^2$: fixing all routing probabilities except $x_k=x$, it has the form $Ax/[1-aB(1-x)]$, with $A=\satDarkFactor\prod_{\ell\neq k}x_\ell$, $a=(1-\darkProb)^2$ and $B=\prod_{\ell\neq k}(1-x_\ell)$, whose derivative $A(1-aB)/[1-aB(1-x)]^2$ is nonnegative; and, for fixed routing probabilities, it has the form $A'/(1-aB')$ with $A',B'\geq0$ and $aB'<1$, which is nondecreasing in $a$. The second inequality holds because the numerator decreases and the denominator increases with $n$, and $\multiCoefficient$ is the value at $n=2$. Note that the entrance transmittance enters this case only through the routing probabilities $x_k$, and the cases (i) and (ii) do not depend on it. Note also that $\multiCoefficient<\satDarkFactor$ for $\maxRouting<1$, since $\maxRouting^2+(1-\darkProb^-)^2(1-\maxRouting)^2\leq\maxRouting^2+(1-\maxRouting)^2<1$.

Since $t\geq0$ and $\satDarkFactor>\multiCoefficient$, the three cases together prove the operator inequality
\begin{equation}\label{eq:operator-bound}
 \inputJointEffect\leq \multiCoefficient \inputClickEffect+(\satDarkFactor-\multiCoefficient)\inputTargetEffect+\vacuumClick (\satDarkFactor-\multiCoefficient)\II_{\earlyInputMode\lateInputMode}.
\end{equation}
Taking the expectation of \cref{eq:operator-bound} with respect to $\density_{\earlyInputMode\lateInputMode}$ and rearranging gives \cref{eq:single_bound}. Note that the argument holds for arbitrary input states, including those with coherences between different photon numbers and correlations with Eve's systems, because the operator inequality itself does not depend on the state. Note also that the coefficients in \cref{eq:operator-bound} are the same for all $\entranceTrans\in[\entranceTrans_-,\entranceTrans_+]$ and all $\darkProb\in[\darkProb^-,\darkProb^+]$, and thus \cref{eq:single_bound} holds if these parameters vary between pulse pairs and between physical windows within their intervals.
\end{proof}

We note that, under \cref{asm:causal}, Eve may use the leakage emitted in the windows $t_1$ and $t_2$ of the current pair to prepare the extra mode $\extraMode{2}$ of the next pair, in which case the target and extra modes are not described by a joint state fixed before the measurements, as assumed above. This, however, does not affect the proof. The reason is that the right-hand side of \cref{eq:extra-domination} acts as the identity on the extra modes, and thus its expectation value does not depend on how, or when, these modes are prepared. In other words, once the no-click events of the satellite windows are bounded as in \cref{eq:extra-domination}, the extra modes play no role in the argument, and \cref{eq:single_bound} follows from the state of the target modes alone, which by \cref{asm:causal} is fixed before the first window involving the pair. We remark that no independence between neighboring pairs is required.

\section{Count estimates with shared satellite windows}
\label[appendix]{app:delayed}
To convert the conditional-probability bounds of the main text into bounds on the observed counts, we apply Azuma's inequality separately to the odd and the even rounds. This grouping is used only in the proof, and both groups contribute to the same key. That is, its purpose is to avoid conditioning a pair on a satellite outcome that already involves that pair, as we explain next.

Let $\fullHistory{i}$ be the increasing classical history containing the choices and measurement records assigned to rounds through $i$, including the late satellite window of round $i$, and let $\fullHistory{i'}$ be the trivial (empty) history for $i'\leq0$. In the chronological realization considered here, all these records are resolved before the first window involving pair $i+2$, and each physical record occurs only once. Current private choices that are deferred in a virtual protocol are included only when sampled there. Unlike $\fullHistory{i-1}$, the history $\fullHistory{i-2}$ used in the main text contains no satellite outcome involving the current target input. The following lemma is the tool that we use to relate conditional probabilities given $\fullHistory{i-2}$ to observed counts.

\begin{Lem}[Delayed concentration]
\label{lem:delayed}
Let $X_1,\dots,X_{\totalPairs}$ be random variables taking values in $[0,1]$ such that each $X_i$ is a function of the history $\fullHistory{i}$. Then, for any $0<\failureProbability<1$, with probability at least $1-\failureProbability$,
\begin{equation}\label{eq:delayed-azuma}
 \left|\sum_{i=1}^{\totalPairs }X_i-\sum_{i=1}^{\totalPairs }\mathbb E[X_i\mid\fullHistory{i-2}]\right|
 \leq2\sqrt{\totalPairs \ln(4/\failureProbability)}.
\end{equation}
\end{Lem}
\begin{proof}
Define
\begin{equation}\label{eq:delayed-difference}
 D_i:=X_i-\mathbb E[X_i\mid\fullHistory{i-2}],
\end{equation}
which satisfies $|D_i|\leq1$ and $\mathbb E[D_i\mid\fullHistory{i-2}]=0$. The odd and even subsequences of $(D_i)_i$ are therefore martingale-difference sequences with respect to the filtrations $(\fullHistory{2k-1})_k$ and $(\fullHistory{2k})_k$, respectively. Note that no independence between the two subsequences, or within each of them, is required. If their sizes are $\totalPairs_o$ and $\totalPairs_e$, a two-sided Azuma bound~\cite{Azuma1967} with failure probability $\failureProbability/2$ for each gives, by a union bound, that with probability at least $1-\failureProbability$ the left-hand side of \cref{eq:delayed-azuma} is at most $\sqrt{2\totalPairs_o\ln(4/\failureProbability)}+\sqrt{2\totalPairs_e\ln(4/\failureProbability)}\leq2\sqrt{\totalPairs\ln(4/\failureProbability)}$, where the last inequality follows from the concavity of the square root and $\totalPairs_o+\totalPairs_e=\totalPairs$. 
\end{proof}
Note that the deviation in~\cref{eq:delayed-azuma} divided by $\totalPairs$ vanishes for $N\to\infty$, and thus $\frac1\totalPairs\sum_iX_i$ and $\frac1\totalPairs\sum_i\mathbb E[X_i\mid\fullHistory{i-2}]$ coincide asymptotically.

We now apply \cref{lem:delayed} to the three estimates of the main text. In what follows, we denote by $\indicator_i^{w,\meanPhoton}$ the indicator that round $i$ has matching basis $w$ and intensity $\meanPhoton$.
For the single-photon estimator, we consider the indicators
\begin{equation}\label{eq:estimator-indicators}
 X_i:=\indicator_i^{w,\meanPhoton}\,\indicator[\mainEvent^{(i)}],\qquad
 \widetilde X_i:=\indicator_i^{w,\meanPhoton}\,\indicator[\mainEvent^{(i)},\satVacEvent^{(i)}],\qquad
 X_i^{1_m}:=\indicator_i^{w,\meanPhoton}\,\indicator[\mainNumber^{(i)}=1],
\end{equation}
whose sums over $i$, divided by $\totalPairs$, are $\observedGain{w}{\meanPhoton}$, $\jointGain_{w,\meanPhoton}$ and $\mainGain^{w,\meanPhoton}$, respectively. Let $p^{(i)}_{w,\meanPhoton}:=\Pr[\indicator_i^{w,\meanPhoton}=1\mid\fullHistory{i-2}]$ denote the conditional probability of the setting. Then $\mathbb E[X_i\mid\fullHistory{i-2}]=p^{(i)}_{w,\meanPhoton}\conditionalMain$, with $\conditionalMain$ the conditional probability of \cref{eq:MTJ}, and similarly for $\widetilde X_i$ and $X_i^{1_m}$. Multiplying \cref{eq:single_bound}, without the truncation $[\cdot]_+$, by $p^{(i)}_{w,\meanPhoton}$ then gives
\begin{equation}\label{eq:estimator-conditional}
 \mathbb E[X_i^{1_m}\mid\fullHistory{i-2}]\geq\frac{\mathbb E[\widetilde X_i\mid\fullHistory{i-2}]-\multiCoefficient\,\mathbb E[X_i\mid\fullHistory{i-2}]}{\satDarkFactor-\multiCoefficient}-p^{(i)}_{w,\meanPhoton}\vacuumClick.
\end{equation}
Summing \cref{eq:estimator-conditional} over the rounds, dividing by $\totalPairs$, and applying \cref{lem:delayed} to the three indicators and to $\indicator_i^{w,\meanPhoton}$---whose sum over $i$, divided by $\totalPairs$, is the observed fraction of rounds with setting $(w,w,\meanPhoton)$, which tends to $\basisProbability{w}^2\intensityProbability$ because the settings are chosen independently in each round---yields \cref{eq:Q1B_lower}, the truncation $[\cdot]_+$ following from $\mainGain^{w,\meanPhoton}\geq0$. We remark that $p^{(i)}_{w,\meanPhoton}$ does not need to equal $\basisProbability{w}^2\intensityProbability$, as Eve may make the records in $\fullHistory{i-2}$ depend on Alice's emitted state in round $i$.

For the phase errors, we use the source-replacement protocol of \cref{subsec:virtual} and consider the indicators $Y_i^{w}$, $Y_i^{\rm ph}$ and $Y_i^{X}$ of a basis-matched target event in basis $w$, of a phase error in a basis-matched key round, and of a bit error in a basis-matched test round, respectively, the last two restricted to the target events. Their sums over $i$, divided by $\totalPairs$, are $\targetGain^{w}$, $\phaseErrorYield^{Z}$ and $\testErrorYield^{X}$. Since Alice's virtual basis and Bob's basis are chosen independently of the state being measured, and the target condition $\mainNumber=1$ is basis independent, we have that
\begin{equation}\label{eq:phase-indicators}
 \mathbb E[Y_i^{w}\mid\fullHistory{i-2}]=\basisProbability{w}^2\operatorname{Tr}\targetState^{(i)},\qquad
 \mathbb E[Y_i^{\rm ph}\mid\fullHistory{i-2}]=\conditionalPhaseError{i},\qquad
 \mathbb E[Y_i^{X}\mid\fullHistory{i-2}]=\conditionalTestError{i},
\end{equation}
and \cref{eq:local-phase}, together with $\phasePenalty\leq\overline{\phasePenalty}$, becomes
\begin{equation}\label{eq:phase-conditional}
 \mathbb E[Y_i^{\rm ph}\mid\fullHistory{i-2}]\leq\frac{\basisProbability{Z}^2}{\basisProbability{X}^2}\,\mathbb E[Y_i^{X}\mid\fullHistory{i-2}]+\overline{\phasePenalty}\,\mathbb E[Y_i^{Z}\mid\fullHistory{i-2}].
\end{equation}
Summing over the rounds and applying \cref{lem:delayed} gives \cref{eq:basis-ratio,eq:phase-yield}, and dividing \cref{eq:phase-yield} by $\targetGain^{Z}$ and using \cref{eq:basis-ratio} gives \cref{eq:phase-rate}. 

Finally, we justify the count identities in \cref{eq:virtual_expansion,eq:error_virtual_expansion}. Let $X_i^{1_m}$ be the indicator defined in \cref{eq:estimator-indicators}, and let $X_i^{1_m,k}$ denote the indicator that round $i$ has matching basis $w$, that Alice emits $k$ photons, and that $\mainNumber^{(i)}=1$, so that $\sum_iX_i^{1_m,k}/\totalPairs=\photonTargetGain{k}^{w}$. Since, at fixed $k$, the emitted state does not depend on the intensity label, and this label is not announced during transmission, the label can be regarded as being sampled only after the detection of round $i$, with the conditional distribution $\posteriorIntensity{k}$ of \cref{eq:decoy-probs}. Hence,
\begin{equation}\label{eq:label-resampling}
 \mathbb E[X_i^{1_m}\mid\fullHistory{i-2}]=\sum_{k\geq0}\posteriorIntensity{k}\,\mathbb E[X_i^{1_m,k}\mid\fullHistory{i-2}],
\end{equation}
and the same relation holds for the indicators of the test-round bit errors with intensity $\meanPhoton$ and with $k$ emitted photons. Summing \cref{eq:label-resampling} over the rounds and applying \cref{lem:delayed} to $X_i^{1_m}$ and to $\sum_{k\geq0}\posteriorIntensity{k}X_i^{1_m,k}$---a random variable determined by $\fullHistory{i}$ and taking values in $[0,1]$, since at most one of the $X_i^{1_m,k}$ is nonzero in each round---gives \cref{eq:virtual_expansion,eq:error_virtual_expansion}.

\section{Extensions for imperfect calibration}
\label[appendix]{app:robust}

In the main text, \cref{thm:single_bound} and the key rate in \cref{eq:key_rate_final} allow the dark-count probability $\darkProb$ and the entrance transmittance $\entranceTrans$ to be known only within calibration intervals, but they require $\darkProb$ to be common to both detectors. Here we show that the analysis can be extended to detectors with unequal, partially characterized dark-count probabilities. Then, we discuss the limits of this extension with respect to the detector efficiency.

\subsection{Unequal and partially characterized dark counts}

Suppose now that detector $\detector{j}$ has independent, input-independent dark-count trials in every main and satellite window, with each probability lying in $[d_j^-,d_j^+]$, where $0\leq d_j^-\leq d_j^+<1$. Let $d_j$ denote its dark-count probability in the current main window.
The single-photon bound of \cref{thm:single_bound} applies unchanged with the coefficients in \cref{eq:q_explicit} replaced by
\begin{equation}\label{eq:dark-tag}
\begin{gathered}
 \vacuumClick:=1-(1-d_0^+)(1-d_1^+),\qquad \satDarkFactor:=u_-^2,\qquad u_-:=(1-d_0^-)(1-d_1^-),\qquad
 \multiCoefficient:=\frac{\satDarkFactor\maxRouting^2}{1-u_-(1-\maxRouting)^2},
\end{gathered}
\end{equation}
where $\vacuumClick$ and $u_-$ upper bound the main-vacuum click probability $1-\darkFreeWeight$ and the probability $\darkFreeWeight:=(1-d_0)(1-d_1)$ of no dark counts in the main window, respectively, and $\satDarkFactor$ upper bounds the probability of no dark counts at either detector in either of the two satellite windows, which is a product of four factors $1-d_j$ with $d_j\geq d_j^-$. Indeed, the proof of \cref{app:proof_lmm_qn} only uses these bounds: it applies directly with $(1-\darkProb)^2$ replaced by $\darkFreeWeight$ in \cref{eq:main-effect,eq:diag-d,eq:diag-f,eq:multi-ratio} and in the three cases, and with $(1-\darkProb^-)^2$ replaced by $u_-$. Hence, the target lower bounds $\targetLower^{w}$ can be computed as in the main text, with the coefficients in \cref{eq:dark-tag}, and the test-error upper bound $\testErrorUpper$ is also unchanged.

The phase-error analysis, however, must be modified. The reason is that, for $\mainNumber=1$, the actual POVM element corresponding to a double click is $d_1\portProjector{0}{w}+d_0\portProjector{1}{w}$, which is not proportional to the identity when $d_0\neq d_1$, and thus the single/double-click branch of \cref{subsec:virtual} can no longer be sampled independently of the input qubit. 

To address this, we decompose the dark counts into a common part, which is treated with the symmetric analysis of the main text, and a residual part, which is present only at the noisier detector and is tagged. Specifically, let $p_c:=\min\{d_0,d_1\}$, and let $\detector{j_*}$ denote the detector with the larger dark-count probability (either of them if $d_0=d_1$). Under the dark-count model of \cref{asm:optical}, the dark-count trial at $\detector{j_*}$ can be realized as the union of two independent trials: a common trial with probability $p_c$, and a residual trial with probability
\begin{equation}\label{eq:residual-dark}
 \zeta:=\frac{\max\{d_0,d_1\}-\min\{d_0,d_1\}}{1-\min\{d_0,d_1\}},
\end{equation}
since $p_c+(1-p_c)\zeta=\max\{d_0,d_1\}$. The residual trial is independent of the input state, of Bob's basis, and of all other systems, and this decomposition is only a relabeling of the dark-count randomness: the recorded pattern, and hence the pattern-conditioned emission of \cref{asm:leak}, are unchanged. Since this is a classical probabilistic mixture generated by Bob's local dark-count randomness, the two branches can be analyzed separately and combined according to their probabilities. Importantly, conditioned on the residual trial not firing, both detectors have independent dark counts with the common probability $p_c$, i.e., the receiver is exactly the symmetric model of the main text with $\darkProb=p_c$. This event has probability
\begin{equation}\label{eq:sym-weight}
 g:=1-\zeta=\frac{1-\max\{d_0,d_1\}}{1-\min\{d_0,d_1\}}
 \geq g_*:=\frac{1-\max\{d_0^+,d_1^+\}}{1-\min\{d_0^-,d_1^-\}}.
\end{equation}
Note that $g_*$ is a uniform lower bound, valid for any $d_0,d_1$ within the calibration intervals, and that it does not require knowing which detector is noisier.

To obtain a good branch $G$ of fixed weight, we use $g\geq g_*$ to decompose the dark-count randomness into a branch $G$ of fixed probability $g_*$, in which the residual trial does not fire, and a residual branch of probability $1-g_*$, in which the residual trial does not fire with probability $(g-g_*)/(1-g_*)$ and fires otherwise. Running the same optical measurement and the same pattern-conditioned emission in both branches reproduces the actual receiver, including its correlations with all other systems, since the overall distribution of the dark-count patterns is unchanged. In the good target events $(1_{\aliceSystem},1_m,G)$, the receiver is the symmetric model with $\darkProb=p_c$, and thus the virtual receiver of \cref{subsec:virtual} applies with the actual value of $p_c$, which the users need not know, with the leakage penalty $\delta=(1-p_c)(1-\leakOverlap)/2\leq\overline{\phasePenalty}:=(1-\min\{d_0^-,d_1^-\})(1-\leakOverlap)/2$ of \cref{eq:delta}, where $\overline{\phasePenalty}$ replaces the bound used in \cref{eq:eph}, since $p_c\geq\min\{d_0^-,d_1^-\}$.

Finally, since the branch $G$ is sampled with fixed probability $g_*$ independently of everything else, the concentration argument of \cref{app:delayed} shows that the number of good target events is asymptotically $g_*$ times the number of target events. Moreover, $\testErrorUpper$ upper bounds all test errors with one emitted photon, and thus also those in the good target events. The corresponding lower bounds and the resulting key rate are therefore
\begin{equation}\label{eq:robust-key}
\begin{aligned}
 \goodTargetLower^{w}&:=g_*\targetLower^{w},\qquad
 \entropyPhase:=\min\left\{\frac{1}{2},\frac{\testErrorUpper}{\goodTargetLower^{X}}+\overline{\phasePenalty}\right\},\\
 \keyRate&=\left[\goodTargetLower^{Z}\big(1-\binaryEntropy (\entropyPhase)\big)-\ecDisclosure\right]_+,
\end{aligned}
\end{equation}
where $\targetLower^{w}$ and $\testErrorUpper$ are computed from \cref{eq:Q1B_lower,eq:Q1_lower_bound}, with the coefficients in \cref{eq:dark-tag}, and \cref{eq:E11_upper_bound}, respectively, and the rate is set to zero if $\goodTargetLower^{Z}\goodTargetLower^{X}=0$. Note that the error-correction term $\ecDisclosure$ is unchanged, because this is a proof-only tag: the users neither identify nor discard physical dark events, and all key rounds are still reconciled. Note also that the tagged fraction $1-g_*$ vanishes as the calibration intervals shrink to a common value, in which case $g_*=1$ and the analysis of the main text is recovered without any tagging. We stress that the bounds $d_j\in[d_j^-,d_j^+]$ must hold in every round, irrespective of the history, rather than only on average; in particular, afterpulsing is not covered by an average dark-count probability, in agreement with \cref{asm:optical}.

\subsection{Efficiency uncertainty and the limits of this extension}

We note that the exact value of the detector efficiency need not be known, as long as it is common to both detectors and independent of the basis, the mode and the window, since it is included in the common loss that we move to the channel. In contrast, overlapping calibration intervals for the two detectors do not establish the equality of their efficiencies, and unequal efficiencies invalidate both the certainty of detection in the single-photon subspace of the main modes---on which \cref{eq:pattern-povm} relies---and, in general, the reduction of the detector loss to an input loss. Accommodating a scalar efficiency mismatch would thus require an additional residual-loss filter or tag, together with the corresponding modifications of the single-photon estimator and of the decoy-state bounds. More generally, spectral, temporal or spatial dependence of the efficiency, detector memory, input-dependent emissions, and arbitrary joint main--satellite leakage are not covered by the formulas proved here. The detector-imperfection framework of Ref.~\cite{Nahar2026ImperfectDetectors} motivates such extensions, but its compatibility with the pattern-conditioned leakage channel of \cref{asm:leak} would also have to be established, and we leave this for future research.

\section{Reproducible loss model for the key-rate curves}
\label[appendix]{sec:sim_setup}

Here we specify the channel model used to generate the observed statistics in \cref{subsec:key-numerics}. We consider a pure-loss channel with transmittance $t=t(d)$, as defined in \cref{sec:sim}, and we assume that neighboring pairs are independently drawn from the same intensity distribution as the current pair. 
For fixed intensity choices, the photon numbers in the modes before the exit beam splitters are independent Poisson variables.

Let us first consider the detection statistics of the main and satellite windows. After the common transmission $t$, the mean photon number in the current main window is $t\meanPhoton/2$, and so its main-vacuum probability is $e^{-t\meanPhoton/2}$. The current pair contributes the same total mean $t\meanPhoton/2$ to its two satellite windows, while the previous late and the next early pulses add the means $t(1-\entranceTrans)\meanPhoton_{i-1}/2$ and $t\entranceTrans\meanPhoton_{i+1}/2$, respectively (see \cref{fig:receiver}). Averaging over the independent neighboring intensity choices gives, for each intensity $\meanPhoton$ of the current pair,
\begin{align}
 A_\entranceTrans(t)&:=\left(\sum_{\meanPhoton'\in\intensitySet}p_{\meanPhoton'} e^{-t(1-\entranceTrans)\meanPhoton'/2}\right)
             \left(\sum_{\meanPhoton'\in\intensitySet}p_{\meanPhoton'} e^{-t\entranceTrans\meanPhoton'/2}\right),\label{eq:neighbor-average}\\
 \conditionalMainMu&=1-(1-\darkProb)^2e^{-t\meanPhoton/2},\qquad
 \conditionalJointMu=\satDarkFactor e^{-t\meanPhoton/2}A_\entranceTrans(t)\conditionalMainMu,\label{eq:channel-MJ}\\
 \observedGain{w}{\meanPhoton}&=\basisProbability{w}^2\intensityProbability \conditionalMainMu,\qquad
 \jointGain_{w,\meanPhoton}=\basisProbability{w}^2\intensityProbability \conditionalJointMu,\label{eq:channel-gains}
\end{align}
%
where $A_\entranceTrans(t)$ is the probability that the neighboring pairs contribute no photon to the two satellite windows, and $\conditionalMainMu$ and $\conditionalJointMu$ are the probabilities $\conditionalMain$ and $\conditionalJoint$ in \cref{eq:MTJ}, averaged over the history at fixed intensity $\meanPhoton$ and matching bases. In \cref{eq:channel-MJ}, we have used that, since both output ports of each satellite window are monitored, its no-click probability does not depend on the interference phase, and that the total photon numbers in the main and satellite windows are independent at fixed $\meanPhoton$.
If all intensities are equal, $A_\entranceTrans=e^{-t\meanPhoton/2}$, recovering \cref{eq:simulation-estimator}.

Next, we consider the bit errors. We choose the port and phase conventions such that detector $b$ is the intended detector for bit $b$ in a matching basis, i.e., the two main-mode amplitudes interfere constructively at $\detector{b}$ when Alice sends bit $b$ and Bob uses the matching basis. The detection probabilities are then the same for the $Z$ and $X$ bases, and the output coherent-state means in a matching basis are
\begin{align}
 \receivedMean_0^{(b)}&=\frac{t\meanPhoton}{2}\big[\exitTrans(1-\entranceTrans)+(1-\exitTrans)\entranceTrans+(-1)^b v\big],\label{eq:port-mean0}\\
 \receivedMean_1^{(b)}&=\frac{t\meanPhoton}{2}\big[(1-\exitTrans)(1-\entranceTrans)+\exitTrans\entranceTrans-(-1)^b v\big],\label{eq:port-mean1}\\
 z_j^{(b)}&:=(1-\darkProb)e^{-\receivedMean_j^{(b)}},\label{eq:port-noclick}
\end{align}
where $v:=2\sqrt{\entranceTrans(1-\entranceTrans)\exitTrans(1-\exitTrans)}$ is the amplitude of the interference term, and $z_j^{(b)}$ is the no-click probability at detector $\detector{j}$ when Alice sends the bit $b$. Note that \cref{eq:port-mean0,eq:port-mean1} are symmetric under $\entranceTrans\leftrightarrow\exitTrans$, as mentioned in \cref{subsec:key-numerics}. In this model, a wrong-detector-only click is an error, and a double click is an error with probability $1/2$. Averaging over equally likely bits therefore gives the error probability per pair at intensity $\meanPhoton$,
\begin{equation}\label{eq:channel-errors}
 B_\meanPhoton:=\frac14\sum_{b=0}^1(1-z_{1-b}^{(b)})(1+z_b^{(b)}),\qquad
 \observedErrors^{X,\meanPhoton}=\basisProbability{X}^2\intensityProbability B_\meanPhoton,\qquad
 \keyBitError=\frac{\sum_\meanPhoton \intensityProbability B_\meanPhoton}{\sum_\meanPhoton \intensityProbability \conditionalMainMu},
\end{equation}
where the last two expressions are the resulting test-error yield and key-basis bit-error rate, respectively. Note that these formulas include the visibility reduction caused by the entrance or exit imbalance, i.e., we do not set the optical error to zero for an unbalanced interferometer. Moreover, in this channel model, the exact gain of the $(1_{\aliceSystem},1_m)$ events is $\targetGain^{w}=\basisProbability{w}^2\emissionProbability{1}t/2$, since a single photon is routed to the main modes with probability $(1-\entranceTrans)/2+\entranceTrans/2=1/2$, which provides a check on the decoy-state lower bound in \cref{eq:Q1_lower_bound}.

We stress that the expressions above are performance inputs only, i.e., they specify the expected values of the observed quantities under this particular channel model. The security proof itself allows arbitrary states and adaptive attacks subject to \cref{asm:optical,asm:causal,asm:leak,asm:source}, and, in particular, it does not require the records of consecutive rounds to be independent.

For the numerical optimization of the key rate, we use $70$ geometrically spaced values of $\meanPhoton_1\in[0.01,1]$ and $50$ geometrically spaced values of $\meanPhoton_2\in[10^{-4},0.1]$, retaining the $3192$ combinations satisfying $\meanPhoton_1>\meanPhoton_2$. At each distance, we maximize the key-rate lower bound over these combinations. The distances are sampled in $5\,\mathrm{km}$ steps, with $0.1\,\mathrm{km}$ steps over the final $5\,\mathrm{km}$ of each positive-rate curve, and the cutoff distance of each curve is located by bisection within a bracket of width below $10^{-6}\,\mathrm{km}$, such that the key rate is positive at its lower end and nonpositive at its upper end. The last plotted point of each curve is placed $0.001\,\mathrm{km}$ below the lower end of this bracket, so that its rate is strictly positive. For $(\entranceTrans,\exitTrans)=(0.5,0.5),(0.4,0.5),(0.7,0.5),(0.5,0.4)$, the cutoff distances are approximately $277.49$, $266.77$, $195.44$, and $266.77\,\mathrm{km}$ at $\leakOverlap=1/\sqrt{2}$, and $289.88$, $285.34$, $268.53$, and $285.34\,\mathrm{km}$ at $\leakOverlap=1$, respectively.

\bibliographystyle{apsrev4-2}
\bibliography{refs_final}
\end{document}

%% file: notation_final.tex
\newcommand{\II}{\mathbb I}
\newcommand{\intensitySet}{\mathcal M}
\newcommand{\entranceTrans}{\eta}
\newcommand{\exitTrans}{\tau}
\newcommand{\meanPhoton}{\mu}
\newcommand{\signalPhase}{\phi}
\newcommand{\globalPhase}{\xi}
\newcommand{\coherentAmplitude}{\alpha}
\newcommand{\receivedMean}{\nu}

\newcommand{\failureProbability}{\epsilon}
\newcommand{\basisProbability}[1]{p_{#1}}
\newcommand{\intensityProbability}{p_{\mu}}
\newcommand{\intensityChoiceProbability}[1]{p_{\mu_{#1}}}
\newcommand{\emissionProbability}[1]{P_{#1}}
\newcommand{\photonProbability}[1]{p_{#1|\mu}}
\newcommand{\posteriorIntensity}[1]{p_{\mu|#1}}
\newcommand{\observedGain}[2]{Q_{#1,#2}}
\newcommand{\totalGain}[1]{Q_{#1}}
\newcommand{\keyBitError}{e_Z}
\newcommand{\mainMode}[1]{m_{#1}}
\newcommand{\satMode}[1]{s_{#1}}
\newcommand{\extraMode}[1]{x_{#1}}
\newcommand{\satRecord}[1]{r_{#1}}
\newcommand{\mainRecord}{r_m}
\newcommand{\alicePhotonNumber}{n_A}
\newcommand{\mainNumberOp}{\hat N_M}
\newcommand{\mainNumber}{N_m}
\newcommand{\mainProjector}[1]{\hat P_M^{#1}}
\newcommand{\numberChannel}{\mathcal D_M}

\newcommand{\qubitSpace}{\mathcal H_B}
\newcommand{\mainSpace}{\mathcal H_M}
\newcommand{\inputSpace}{\mathcal H_{el}}
\newcommand{\fullHistory}[1]{\mathcal F_{#1}} 
\newcommand{\mainEvent}{\mathbf1_{\rm main}}
\newcommand{\satVacEvent}{\mathbf0_{\rm sat}}
\newcommand{\indicator}{\mathbf1}
\newcommand{\detEfficiency}{\eta_{\rm det}}
\newcommand{\darkProb}{p_d}
\newcommand{\leakOverlap}{\gamma}

\newcommand{\phasePenalty}{\delta}
\newcommand{\ecDisclosure}{\lambda_{\rm EC}}
\newcommand{\ecInefficiency}{f_{\rm EC}}
\newcommand{\keyRate}{R_{\rm lower}}
\newcommand{\satDarkFactor}{q_0}
\newcommand{\multiCoefficient}{q_2}
\newcommand{\bobPhase}{\varphi}
\newcommand{\entranceIsometry}{{\hat V^{\entranceTrans}_{\earlyInputMode\lateInputMode\to\mainSystem\satelliteSystem}}} 
\newcommand{\portProjector}[2]{\hat\Pi_B^{#1,#2}}

\newcommand{\testErrorEffect}{\hat E_{AB}^{X}}
\newcommand{\phaseErrorEffect}{\hat E_{AB}^{\rm ph}}
\newcommand{\mainClickEffect}{\hat D_M}
\newcommand{\inputClickEffect}{\hat D_{el}}
\newcommand{\inputTargetEffect}{\hat T_{el}^{m}}
\newcommand{\inputJointEffect}{\hat F_{el}^{0}}
\newcommand{\vacuumProjector}[1]{\hat P_{#1}^{\rm vac}}
\newcommand{\density}{\hat\rho}
\newcommand{\leakState}{\hat\sigma}
\newcommand{\targetState}{\widetilde{\rho}_{AB}}

\newcommand{\goodTargetLower}{\underline Q_{1_A,1_m,G}}
\newcommand{\targetLower}{\underline Q_{1_A,1_m}}
\newcommand{\mainLower}{\underline Q_{1_m}}
\newcommand{\testErrorUpper}{\overline E_{1_A,1_m}^{X}}
\newcommand{\entropyPhase}{\bar e_{\rm ph}}
\newcommand{\targetGain}{Q_{1_A,1_m}}
\newcommand{\photonTargetGain}[1]{Q_{#1_A,1_m}}
\newcommand{\mainGain}{Q_{1_m}}
\newcommand{\phaseErrorYield}{E_{{\rm ph},1_A,1_m}}
\newcommand{\testErrorYield}{E_{1_A,1_m}}
\newcommand{\targetPhaseRate}{e_{{\rm ph},1_A,1_m}}
\newcommand{\targetTestRate}{e_{1_A,1_m}}
\newcommand{\observedErrors}{E_{\rm obs}}
\newcommand{\jointGain}{\widetilde Q}
\newcommand{\totalPairs}{N}
\newcommand{\vacuumClick}{c}
\newcommand{\maxRouting}{h}
\newcommand{\conditionalMain}{M}
\newcommand{\conditionalJoint}{J}
\newcommand{\conditionalTarget}{T}
\newcommand{\binaryEntropy}{H}
\newcommand{\virtualBranch}{R}
\newcommand{\darkFreeWeight}{r}
\newcommand{\numberOperator}{\hat n}
\newcommand{\sourceKetSymbol}{\Psi}
\newcommand{\replacementKetSymbol}{\Omega}
\newcommand{\earlySignalMode}{E}
\newcommand{\lateSignalMode}{L}
\newcommand{\signalSystem}{S}
\newcommand{\earlyInputMode}{e}
\newcommand{\lateInputMode}{l}
\newcommand{\mainSystem}{M}
\newcommand{\satelliteSystem}{S_{\rm sat}}
\newcommand{\aliceSystem}{A}
\newcommand{\bobSystem}{B}

\newcommand{\purificationSystem}{F}
\newcommand{\mainLeakageSystem}{L_m}
\newcommand{\satLeakageSystem}[1]{L_{#1}}
\newcommand{\detector}[1]{D_{#1}}
\newcommand{\alicePhotonErrors}[1]{E_{#1_A}}
\newcommand{\conditionalPhaseError}[1]{p_{{\rm ph},#1}}
\newcommand{\conditionalTestError}[1]{p_{X,{\rm err},#1}}

%% file: refs_final.bib
@article{Gottesman2004ImperfectDevices,
  author  = {Daniel Gottesman and Hoi-Kwong Lo and Norbert L{\"u}tkenhaus and John Preskill},
  title   = {Security of quantum key distribution with imperfect devices},
  journal = {Quantum Information and Computation},
  volume  = {4},
  number  = {5},
  pages   = {325--360},
  year    = {2004},
  doi     = {10.26421/QIC4.5-1}
}

@article{Tamaki2014LossTolerant,
  author  = {Kiyoshi Tamaki and Marcos Curty and Go Kato and Hoi-Kwong Lo and Koji Azuma},
  title   = {Loss-tolerant quantum cryptography with imperfect sources},
  journal = {Physical Review A},
  volume  = {90},
  pages   = {052314},
  year    = {2014},
  doi     = {10.1103/PhysRevA.90.052314}
}

@article{Pereira2020CorrelatedSources,
  author  = {Margarida Pereira and Go Kato and Akihiro Mizutani and Marcos Curty and Kiyoshi Tamaki},
  title   = {Quantum key distribution with correlated sources},
  journal = {Science Advances},
  volume  = {6},
  number  = {37},
  pages   = {eaaz4487},
  year    = {2020},
  doi     = {10.1126/sciadv.aaz4487}
}

@article{CurrasLorenzo2025ImperfectSources,
  author  = {Guillermo Curr{\'a}s-Lorenzo and Margarida Pereira and Go Kato and Marcos Curty and Kiyoshi Tamaki},
  title   = {Security framework for quantum key distribution with imperfect sources},
  journal = {Optica Quantum},
  volume  = {3},
  number  = {6},
  pages   = {525--534},
  year    = {2025},
  doi     = {10.1364/OPTICAQ.569424}
}

@misc{Navarrete2026NumericalPracticalQKD,
  author        = {{\'A}lvaro Navarrete and Guillermo Curr{\'a}s-Lorenzo and Margarida Pereira and Marcos Curty},
  title         = {Numerical security analysis for practical quantum key distribution},
  year          = {2026},
  eprint        = {2605.12984},
  archivePrefix = {arXiv},
  primaryClass  = {quant-ph}
}

@article{Lo2012MDIQKD,
  author  = {Hoi-Kwong Lo and Marcos Curty and Bing Qi},
  title   = {Measurement-device-independent quantum key distribution},
  journal = {Physical Review Letters},
  volume  = {108},
  pages   = {130503},
  year    = {2012},
  doi     = {10.1103/PhysRevLett.108.130503}
}

@article{Li2020FlagStateSquashing,
  author  = {Nicky Kai Hong Li and Norbert L{\"u}tkenhaus},
  title   = {Improving key rates of the unbalanced phase-encoded {BB84} protocol using the flag-state squashing model},
  journal = {Physical Review Research},
  volume  = {2},
  pages   = {043172},
  year    = {2020},
  doi     = {10.1103/PhysRevResearch.2.043172}
}

@article{Zhang2021DetectionEfficiencyMismatch,
  author  = {Yanbao Zhang and Patrick J. Coles and Adam Winick and Jie Lin and Norbert L{\"u}tkenhaus},
  title   = {Security proof of practical quantum key distribution with detection-efficiency mismatch},
  journal = {Physical Review Research},
  volume  = {3},
  pages   = {013076},
  year    = {2021},
  doi     = {10.1103/PhysRevResearch.3.013076}
}

@article{Kamin2024ImprovedFlagState,
  author  = {Lars Kamin and Norbert L{\"u}tkenhaus},
  title   = {Improved decoy-state and flag-state squashing methods},
  journal = {Physical Review Research},
  volume  = {6},
  pages   = {043223},
  year    = {2024},
  doi     = {10.1103/PhysRevResearch.6.043223}
}

@article{TsurumaruTamaki2008ThresholdDetectors,
  author  = {Toyohiro Tsurumaru and Kiyoshi Tamaki},
  title   = {Security proof for quantum-key-distribution systems with threshold detectors},
  journal = {Physical Review A},
  volume  = {78},
  pages   = {032302},
  year    = {2008},
  doi     = {10.1103/PhysRevA.78.032302},
  eprint  = {0803.4226},
  archivePrefix = {arXiv},
  primaryClass  = {quant-ph}
}

@article{Beaudry2008Squashing,
  author  = {Normand J. Beaudry and Tobias Moroder and Norbert L{\"u}tkenhaus},
  title   = {Squashing Models for Optical Measurements in Quantum Communication},
  journal = {Physical Review Letters},
  volume  = {101},
  pages   = {093601},
  year    = {2008},
  doi     = {10.1103/PhysRevLett.101.093601},
  eprint  = {0804.3082},
  archivePrefix = {arXiv},
  primaryClass  = {quant-ph}
}

@article{Gittsovich2014Squashing,
  author  = {Oleg Gittsovich and Normand J. Beaudry and Varun Narasimhachar and Ruben Romero Alvarez and Tobias Moroder and Norbert L{\"u}tkenhaus},
  title   = {Squashing model for detectors and applications to quantum-key-distribution protocols},
  journal = {Physical Review A},
  volume  = {89},
  pages   = {012325},
  year    = {2014},
  doi     = {10.1103/PhysRevA.89.012325},
  eprint  = {1310.5059},
  archivePrefix = {arXiv},
  primaryClass  = {quant-ph}
}

@article{Nahar2026ImperfectDetectors,
  author = {Shlok Nahar and Devashish Tupkary and Norbert L{\"u}tkenhaus},
  title = {Imperfect detectors for adversarial tasks with applications to quantum key distribution},
  journal = {Quantum},
  volume = {10},
  pages = {2044},
  year = {2026},
  doi = {10.22331/q-2026-03-24-2044},
  eprint = {2503.06328},
  archivePrefix = {arXiv}
}

@article{charles2013,
  author  = {Lim, C. C. W. and Curty, M. and Walenta, N. and Xu, F. and Zbinden, H.},
  title   = {Concise security bounds for practical decoy-state quantum key distribution},
  journal = {Physical Review A},
  volume  = {89},
  pages   = {022307},
  year    = {2014}
}

@article{leakyreceiver2026,
  author        = {Navarrete, {\'A}. and Pereira, M. and Curr{\'a}s-Lorenzo, G. and Mizutani, A. and Curty, M. and Tamaki, K.},
  title         = {Phase-error estimation for quantum key distribution with leaky receivers},
  journal       = {arXiv preprint arXiv:2608.09674},
  year          = {2026}
}

@article{tomamichel2011,
  author  = {Tomamichel, M. and Renner, R.},
  title   = {The uncertainty relation for smooth entropies},
  journal = {Physical Review Letters},
  volume  = {106},
  pages   = {110506},
  year    = {2011}
}

@article{tomamichel2017,
  author  = {Tomamichel, M. and Leverrier, A.},
  title   = {A largely self-contained and complete security proof for quantum key distribution},
  journal = {Quantum},
  volume  = {1},
  pages   = {14},
  year    = {2017}
}

@article{moroder2009detector,
  title     = {Detector decoy quantum key distribution},
  author    = {Moroder, Tobias and Curty, Marcos and L{\"u}tkenhaus, Norbert},
  journal   = {New Journal of Physics},
  volume    = {11},
  number    = {4},
  pages     = {045008},
  year      = {2009},
  publisher = {IOP Publishing},
  doi       = {10.1088/1367-2630/11/4/045008}
}

@article{tomamichel2011hashing,
  author    = {Tomamichel, Marco and Schaffner, Christian and Smith, Adam and Renner, Renato},
  title     = {Leftover Hashing Against Quantum Side Information},
  journal   = {IEEE Transactions on Information Theory},
  volume    = {57},
  number    = {8},
  pages     = {5524--5535},
  year      = {2011},
  publisher = {IEEE}
}

@article{Azuma1967,
  author  = {Kazuoki Azuma},
  title   = {Weighted sums of certain dependent random variables},
  journal = {Tohoku Mathematical Journal, Second Series},
  volume  = {19},
  number  = {3},
  pages   = {357--367},
  year    = {1967},
  doi     = {10.2748/tmj/1178243286},
  url     = {https://www.jstage.jst.go.jp/article/tmj1949/19/3/19_3_357/_article/-char/en}
}

@inproceedings{bb84,
  author    = {Bennett, Charles H. and Brassard, Gilles},
  title     = {Quantum cryptography: Public key distribution and coin tossing},
  booktitle = {Proc. IEEE Int. Conf. Computers, Systems and Signal Processing},
  pages     = {175--179},
  year      = {1984},
  address   = {Bangalore, India}
}

@article{lo2014secure,
  author  = {Lo, Hoi-Kwong and Curty, Marcos and Tamaki, Kiyoshi},
  title   = {Secure quantum key distribution},
  journal = {Nat. Photonics},
  volume  = {8},
  pages   = {595--604},
  year    = {2014}
}

@article{pirandola2020advances,
  author  = {Pirandola, Stefano and Andersen, Ulrik L. and Banchi, Leonardo and Berta, Mario and Bunandar, Darius and Colbeck, Roger and Englund, Dirk and Gehring, Tobias and Lupo, Cosmo and Ottaviani, Carlo and others},
  title   = {Advances in quantum cryptography},
  journal = {Adv. Opt. Photonics},
  volume  = {12},
  pages   = {1012--1236},
  year    = {2020}
}

@article{xu2020secure,
  author  = {Xu, Feihu and Ma, Xiongfeng and Zhang, Qiang and Lo, Hoi-Kwong and Pan, Jian-Wei},
  title   = {Secure quantum key distribution with realistic devices},
  journal = {Rev. Mod. Phys.},
  volume  = {92},
  pages   = {025002},
  year    = {2020}
}

@techreport{marquardt2024implementation,
  author      = {Marquardt, Christoph and Seyfarth, Ulrich and Bettendorf, Sven and Bohmann, Martin and Buchner, Alexander and Curty, Marcos and Elser, Dominique and Eul, Silas and Gehring, Tobias and Jain, Nitin and Klocke, Thomas and Reinecke, Marie and Sieber, Nico and Ursin, Rupert and Wehling, Marc and Weier, Henning},
  title       = {Implementation attacks against {QKD} systems},
  institution = {German Federal Office for Information Security (BSI)},
  year        = {2024},
  note        = {\url{https://www.bsi.bund.de/EN/Service-Navi/Publikationen/Studien/QKD-Systems/Implementation_Attacks_QKD_Systems_node.html}}
}

@article{zapatero2025,
  author  = {Zapatero, V{\'i}ctor and Navarrete, {\'A}lvaro and Curty, Marcos},
  title   = {Implementation security in quantum key distribution},
  journal = {Adv. Quantum Technol.},
  volume  = {8},
  pages   = {2300380},
  year    = {2025}
}

@article{wiesemann2026security,
  title={Security framework for practical quantum key distribution with imperfect devices},
  author={Wiesemann, Jerome and Burniston, John and Tupkary, Devashish and L{\"u}tkenhaus, Norbert},
  journal={arXiv preprint arXiv:2609.15790},
  year={2026}
}

@article{curras2026phase,
  title={Phase-error-estimation security framework for quantum key distribution with correlated sources},
  author={Curr{\'a}s-Lorenzo, Guillermo and Pereira, Margarida and Tamaki, Kiyoshi and Curty, Marcos},
  journal={Physical Review Research},
  volume={8},
  number={3},
  pages={L032045},
  year={2026},
  publisher={APS}
}

@article{curras2025numerical,
  title={Numerical security analysis for quantum key distribution with partial state characterization},
  author={Curr{\'a}s-Lorenzo, Guillermo and Navarrete, {\'A}lvaro and N{\'u}{\~n}ez-Bon, Javier and Pereira, Margarida and Curty, Marcos},
  journal={Quantum Science and Technology},
  volume={10},
  number={3},
  pages={035031},
  year={2025},
  publisher={IOP Publishing}
}

@article{Lo2005DecoyState,
  author  = {Hoi-Kwong Lo and Xiongfeng Ma and Kai Chen},
  title   = {Decoy State Quantum Key Distribution},
  journal = {Phys. Rev. Lett.},
  volume  = {94},
  pages   = {230504},
  year    = {2005},
  doi     = {10.1103/PhysRevLett.94.230504}
}

@article{CurrasLorenzo2026SourceDetector,
  author  = {Guillermo Curr{\'a}s-Lorenzo and Margarida Pereira and Shlok Nahar and Devashish Tupkary},
  title   = {Security of quantum key distribution with source and detector imperfections through phase-error estimation},
  journal = {npj Quantum Information},
  year    = {2026},
  doi     = {10.1038/s41534-026-01264-5},
  eprint  = {2507.03549},
  archivePrefix = {arXiv},
  primaryClass  = {quant-ph}
}

@article{tupkary2025phase,
  title={Phase error rate estimation in QKD with imperfect detectors},
  author={Tupkary, Devashish and Nahar, Shlok and Sinha, Pulkit and L{\"u}tkenhaus, Norbert},
  journal={Quantum},
  volume={9},
  pages={1937},
  year={2025},
  publisher={Verein zur F{\"o}rderung des Open Access Publizierens in den Quantenwissenschaften}
}

@article{Kurtsiefer2001Backflash,
  author  = {Kurtsiefer, Christian and Zarda, Patrick and Mayer, Sonja and Weinfurter, Harald},
  title   = {The breakdown flash of silicon avalanche photodiodes---back door for eavesdropper attacks?},
  journal = {J. Mod. Opt.}, volume = {48}, number = {13}, pages = {2039--2047}, year = {2001}}

@article{Meda2017Backflash,
  author  = {Meda, Alice and Degiovanni, Ivo P. and Tosi, Alberto and Yuan, Zhiliang and Brida, Giorgio and Genovese, Marco},
  title   = {Quantifying backflash radiation to prevent zero-error attacks in quantum key distribution},
  journal = {Light Sci. Appl.}, volume = {6}, pages = {e16261}, year = {2017}}

@article{Pinheiro2018Backflash,
  author  = {Pinheiro, Paulo Vinicius Pereira and Chaiwongkhot, Poompong and Sajeed, Shihan and Horn, Rolf T. and Bourgoin, Jean-Philippe and Jennewein, Thomas and L{\"u}tkenhaus, Norbert and Makarov, Vadim},
  title   = {Eavesdropping and countermeasures for backflash side channel in quantum cryptography},
  journal = {Opt. Express}, volume = {26}, number = {16}, pages = {21020--21032}, year = {2018}}

@article{maroySecurityQuantum2010,
  author  = {Mar{\o}y, {\O}ystein and Lydersen, Lars and Skaar, Johannes},
  title   = {Security of quantum key distribution with arbitrary individual imperfections},
  journal = {Phys. Rev. A},
  volume  = {82},
  pages   = {032337},
  year    = {2010},
  doi     = {10.1103/PhysRevA.82.032337}
}

@misc{arqandMutualInformation2024,
  author        = {Arqand, Amir and Metger, Tony and Tan, Ernest Y.-Z.},
  title         = {Mutual information chain rules for security proofs robust against device imperfections},
  year          = {2024},
  eprint        = {2407.20396},
  archivePrefix = {arXiv},
  primaryClass  = {quant-ph}
}
